\documentclass[11pt]{article}
\usepackage[latin9]{inputenc}
\usepackage{tikzit}
\usepackage{geometry}
\usepackage{color}
\usepackage{xcolor}

\ifx11  % Use '11' to include colors and 10 to exclude them
\definecolor{darkorange}{rgb}{1.0, 0.55, 0.0}
\definecolor{olive}{rgb}{0.5, 0.5, 0.0}
\newcommand{\oliveq}{olive!15}
\newcommand{\darkorangeq}{darkorange!15}
\definecolor{darkgreen}{rgb}{0,0.7,0}
\else
\definecolor{darkorange}{rgb}{0,0,0}
\definecolor{olive}{rgb}{0,0, 0.0}
\definecolor{darkgreen}{rgb}{0,0.,0}
\newcommand{\oliveq}{white}
\newcommand{\darkorangeq}{white}
\fi

\definecolor{ao(english)}{rgb}{0.0, 0.6, 0.0}
\newcommand{\green}[1]{\textcolor{ao(english)}{#1}}
\usepackage{mathtools}

\DeclarePairedDelimiter\floor{\lfloor}{\rfloor}
\usepackage{amsmath}
\usepackage{amsthm}
\usepackage{amssymb}
\usepackage{graphicx}
\usepackage[authoryear]{natbib}
\usepackage{float} %enables H option for floats
\usepackage{booktabs,makecell} % nicer tables (float for H position)
\usepackage{caption} % nicer caption
\usepackage[labelformat=simple]{subcaption}

\usepackage[unicode=true,pdfusetitle,
bookmarks=true,bookmarksnumbered=true,bookmarksopen=true,bookmarksopenlevel=2,
 breaklinks=true,pdfborder={0 0 1},backref=page,colorlinks=true]
 {hyperref}
\hypersetup{
 breaklinks,pdfborderstyle=,allcolors=darkblue}

\makeatletter

\usepackage{thmtools}

\declaretheorem [name = Lemma]{lem}

\declaretheorem [name = Proposition]{prop}

\usepackage{multirow}
\usepackage{subcaption}
\usepackage{bbm}
\usepackage{amsfonts}
\usepackage{mathrsfs}
\newcommand{\x}{\boxed}
\renewcommand{\qedsymbol}{$\blacksquare$}

\usepackage{enumerate}
\usepackage[shortlabels]{enumitem}
\usepackage{thmtools}

\usepackage{xspace} %this is for taking care of spacing after the newcommands below
\newcommand{\STR}{\textit{STR}\xspace}
\newcommand{\SAR}{\textit{REP}\xspace}
\newcommand{\RAN}{\textit{RAN}\xspace}
\newcommand{\ph}{{\text -}}
\usepackage[onehalfspacing]{setspace}
\definecolor{darkblue}{rgb}{0,0.1,0.5}
\newcommand{\p}{\mathbb{P}}

\newcommand{\T}{\mathbb{T}}
\newcommand{\M}{\mathbb{A}}
\newcommand{\ul}{\underline}
\newcommand{\ra}{\rightarrow}
\newcommand{\ol}{\overline}

\makeatother

\begin{document}
\singlespacing
\title{Exclusion of Extreme Jurors and Minority Representation: The Effect of Jury Selection Procedures\thanks{Moro: Vanderbilt University, \texttt{andrea@andreamoro.net}. Van Der
Linden: Emory University \texttt{martin.van.der.linden@emory.edu}. We thank Alberto Bisin, Matias Iaryczower, Nicola Persico, and one anonymous referee for helpful comments and suggestions.
We are also grateful to seminar audiences at North Carolina State University, the Discrimination and Disparities seminar, the NBER Political Economy Program Summer Institute, and Iowa State University.
} 
}
\author{Andrea Moro and Martin Van der Linden}
\date{\today \\ 
%{\small First version: February 12, 2021}
}
\maketitle
\begin{abstract} 
    \noindent     We compare two jury selection procedures meant to safeguard against the inclusion of biased jurors that are perceived as causing minorities to be under-represented. 
    The Strike and Replace procedure presents potential jurors one-by-one to the parties, while the Struck procedure presents all potential jurors before the parties exercise their challenges. 
    Struck more effectively excludes extreme jurors but leads to a worse representation of minorities.
    The advantage of Struck in terms of excluding extremes is sizable in a wide range of cases. 
    In contrast, Strike and Replace better represents minorities only if the minority and majority are polarized. Results are robust to assuming the parties statistically discriminate against jurors based on group identity.
\end{abstract}
\vspace{2em}



\thispagestyle{empty}
\newpage \setcounter{page}{1}
\setstretch{1.4} % double spacing too much, 1.5 too little
\section{Introduction}
In the United States legal system, it is customary to let the parties involved in a jury trial dismiss some of the potential jurors without justification. 
These dismissals, known as \emph{peremptory challenges}, are meant to enable ``each side to exclude those jurors it believes will be most partial toward the other side" thereby ``eliminat[ing] extremes of partiality on both sides".\footnote{\label{foot:partial}\emph{Holland v. Illinois}, 493 U.S. 474, 484 (1990).
}
In the last decades, however, peremptory challenges have often been criticized, mainly because they are perceived as causing some groups --- in particular minorities --- to be under-represented in juries.\footnote{\label{foot:Batson}%
	For examples of this line of argument against peremptory challenges, see \cite{sacksChallengingPeremptoryChallenge1989}, \cite{broderick_why_1992},  \cite{hochmanAbolishingPeremptoryChallenge1993}, \cite{marderGenderPeremptoryChallenges1994}, and \cite{smithCallAbolishPeremptory2014}.
	Despite these attacks, the U.S. has so far resisted abandoning peremptory challenges altogether (unlike other countries; like the U.K., where they were abolished in 1988).
	Peremptory challenges remain pervasive in the U.S. and have been affirmed by the U.S. Supreme Court as ``one of the most important rights secured to the accused" \citep[\emph{Swain v. Alabama} 380 U.S. 202 (1965), see][]{_criminal_2009}.
}

The procedure used to let the parties exercise their challenges varies greatly across jurisdictions and is sometimes left to the discretion of the judge.\footnote{\label{foot:flexible}%
	For example, in criminal cases in Illinois, ``[State Supreme Court] Rule 434(a) expressly grants a trial court the discretion to alter the traditional procedure for impaneling juries so long as the parties have adequate notice of the system to be used and the method does not unduly restrict the use of peremptory challenges" (\emph{People v. McCormick}, 328 Ill.App.3d 378, 766 N.E.2d 671, (2d Dist., 2002)).
}
Two classes of procedures are most frequently used.
In Struck procedures (henceforth: \STR), the parties can observe and extensively question \emph{all} the jurors who could potentially serve on their trial \emph{before} exercising their challenges (this questioning process is known as \emph{voir dire}).
In contrast, in Strike and Replace procedures (henceforth: \SAR), smaller groups of jurors are sequentially presented to the parties.
The parties observe and question the group they are presented with (sometimes a single juror) but must exercise their challenges on that group \emph{without} knowing the identity of the next potential jurors.

The goal of this paper is to shed light on the debate that emerged in the legal doctrine over the relative effectiveness of \STR and \SAR at satisfying the two objectives of excluding extreme jurors and ensuring adequate group representation. 
\citet[][pp. 93-94]{bermant_voir_1981}, for example, argues that, by avoiding uncertainty, \STR ``always gives advocates more information on which to base their challenges, and, therefore, [...] is always to be preferred".
Bermant further notes that ``a primary purpose of peremptory challenges is to eliminate extremes of partiality on both sides" and that ``the superiority of the struck jury method in accomplishing this purpose is manifest."

Others have argued that, by revealing the identity of all potential jurors before challenges are exercised, \STR facilitates the exclusion of some groups from juries.
In \emph{Batson v. Kentucky}, and \emph{J.E.B. v. Alabama} the Supreme Court found it unconstitutional to challenge potential jurors based on their race or gender.\footnote{476 U.S. 79 (1986); see also \emph{J. E. B. v. Alabama}, 511 U.S. 127 (1994).
The response to these decisions has consisted in allowing the parties to appeal peremptories from their opponent, so that peremptories proven to be based merely on the juror's race can be nullified. 
These appeals are known as \emph{Batson appeals}.} 
However, proving that a challenge is based on race or gender is often difficult, and the Supreme Court's ruling is therefore notoriously hard to enforce.\footnote{
	See \citet{raphael_excuses_1993}: ``In virtually any situation, an intelligent plaintiff can produce a plausible neutral explanation for striking [a black juror] despite the plaintiff's having acted on racial bias.
	Consequently, given the current case law, a plaintiff who wishes to offer a pretext for a race-based strike is unlikely to encounter difficulty in crafting a neutral explanation.'' See also \citet{marderBatsonRevisitedBatson2012}
	or \citet{dalyFosterChatmanClarifying2016} for why judges rarely
	rule in favor of Batson appeals. } 
Interestingly, in response, judges themselves have turned to the design of the challenge procedure and the use of \SAR as an instrument to foster adequate group representation.
In a memorandum on judges' practices regarding jury selection, \citet{_cr51995pdf-1} for example report about judges believing that by ``prevent[ing] counsel from knowing who might replace a challenged juror" \SAR procedures ``make it more difficult to pursue a strategy prohibited by \emph{Batson}."\footnote{
Some have gone further and argued for removing peremptory challenges altogether as a more drastic protection against the exclusion of jurors by race.
In August 2021 the Supreme Court of Arizona ordered the elimination of challenges altogether (Arizona Supreme Court No. R-21-0020, available at \href{https://www.azcourts.gov/Rules-Forum/aft/1208}{https://www.azcourts.gov/Rules-Forum/aft/1208}).}

To inform this debate, we extend in Section \ref{sec:model} the model of jury selection proposed in \citet{brams_optimal_1978} by allowing potential jurors to belong to two different groups. 
In the model, each potential juror is characterized by a probability to vote in favor of the defendant's conviction.
This probability is drawn from a distribution that depends on the juror's group-membership. 
The group distributions are common knowledge but the parties to the trial, a plaintiff and a defendant, only observe their realization for a particular juror upon questioning that juror.

A jury must be formed to decide the outcome of the trial and the parties can influence its composition by challenging (i.e., vetoing) a certain number of potential jurors.
Challenges are exercised according to \SAR or \STR procedures which, as explained above, differ mainly in the timing of jurors' questioning (and, as a consequence, in the parties' ability to observe the conviction probability of potential jurors).

We ask how these two procedures perform in achieving the objectives of excluding extreme jurors and ensuring adequate group representation. 
In Section \ref{sec:example}, we introduce an illustrative example where a single juror must be selected, and the parties each have a single challenge available. 
In this example, we show that \STR is more effective than \SAR at excluding jurors from the tails of the conviction probability distribution, but is less likely to select minority jurors.

The rest of the paper is devoted to characterizing conditions under which these results extend beyond the illustrative example of Section \ref{sec:example}.
In Section \ref{sec:extreme} we call a juror \emph{extreme} if its conviction probability falls below (above) a given threshold. 
We prove that there always exists a low enough threshold such that \STR is more likely than \SAR to exclude extreme jurors. 
Moreover, we show that \STR always selects fewer extreme jurors than a random selection would, but that there are some (admittedly somewhat unusual) circumstances in which \SAR would not. 
Simulations assuming a wide range of conviction probability distributions reveal that, in terms of excluding extreme jurors, the advantage of \STR over \SAR can be substantial, even for relatively high thresholds.

Section \ref{sec:grouprep} compares procedures according to their ability to select minorities, identifying conditions under which \SAR selects more minority jurors than \STR. 
Our proof uses a limiting argument showing that the result holds when the minority is vanishingly small and the distributions of conviction probabilities for each group minimally overlap (i.e., groups are polarized).
However, simulations suggest that the result remains true when the size of the minority is relatively high and the overlap between distributions is significant. 

In Section \ref{sec:number}, we explore how changing the number of challenges affects the results of Sections \ref{sec:extreme} and \ref{sec:grouprep}. In any procedure, increasing the number of challenges helps the exclusion of more extreme jurors, but reduces minority representation. In many jurisdictions, more challenges are granted when the charge is more severe, such as in capital cases.\footnote{See \cite{rottman2004dep}} With a minority defendant, if minority jurors are less likely to vote conviction, these norms favor the prosecution. 

In Section \ref{sec:discrimination} we show that the results we obtained extend to a setup in which parties observe conviction probability with noise, and rely on noisy signals and group identity to make their choices, a form of statistical discrimination.

Finally, in Section \ref{sec:ext} we show how our main theoretical results extend to a different definition of extreme juries (i.e., a jury in which the \emph{highest (lowest)} conviction-probability juror is below (above) a given threshold).

We also explore how the procedures compare in selecting members of groups that are of similar sizes (such as males and females, as opposed to minorities which induce groups of unequal sizes). 

\subsection*{Related Literature} 
This paper belongs to a relatively small literature formalizing jury selection procedures. 
\cite{brams_optimal_1978} model \SAR as a game and derive its subgame-perfect equilibrium strategies which we use in our theoretical results and simulations. 
Perhaps closest to our paper is \cite{flanagan_peremptory_2015} who shows that, compared to randomly selecting jurors, \STR increases the probability that all jurors come from one particular side of the median of the distribution of conviction probabilities (because \STR induces correlation between the conviction probability of the selected jurors).
To our knowledge, this literature is silent on the implications of jury  selection for group representation and on the trade-off between excluding extreme jurors and ensuring adequate group representation induced by using different procedures.
These implications are the focus of this paper.

While the group composition of a jury has been shown to influence the outcome of a trial \citep{anwar_impact_2012, anwar2019jury, anwar22, flanaganRaceGenderJuries2018, hoekstra2021}, legal scholars often argue in favor of representative juries regardless of their effect on verdicts.\footnote{Using jury data from Texas, \cite{anwar22} show that another important element affecting outcomes is the selection of the jury pool, which we ignore in this paper.}
\cite{diamondAchievingDiversityJury2009} for example argue that ``unrepresentative juries [...] threaten the public's faith in the legitimacy of the legal system."
In an experiment on jury-eligible individuals, they show that participants rate the outcome of trials as significantly fairer when the jury is racially heterogeneous than when it is not.
This motivates us to consider group-representativity itself as a desirable feature of  jury selection procedures.

% \footnote{
% 	One might also be interested in the impact of group-representation on the conviction of defendants who themselves belong to different groups.
% 	Without taking groups into account or attempting to compare procedures, \cite{flanagan_peremptory_2015} studies the impact of  jury selection procedures on conviction rates.
% 	His results in terms of conviction rates require to assume that the parties have correct beliefs about the probability that jurors eventually vote for conviction (as well as about these probabilities are independent of one another).
% 	In contrast, our results about group-representation and exclusion of extremes do not require that the parties' belief at the moment of  jury selection be accurate (at least if we are concerned with extremes \emph{as perceived by the parties}, as the U.S. Supreme Court seems to be when saying that the main purpose of peremptory challenges is to enable ``each side to exclude those jurors \emph{it believes} will be most partial toward the other side", see Footnote \ref{foot:partial} and associated quote).\martinLi{Note to self: Probably too long and too detailed. Revise.} \andreai{Begs the question of why we don't study outcomes. We could add a para saying we don't do it cause we don't want to take a stand on aggregation of preferences}
% }

The empirical literature on jury selection has also identified systematic patterns of group-specific challenges from the parties, with the plaintiffs being almost always more likely to remove minority jurors than defendants \citep{anwar2014role, anwar22, willcraft2028, diamondAchievingDiversityJury2009,  flanaganRaceGenderJuries2018, rosePeremptoryChallengeAccused1999a,turnerRacePeremptoryChallenges1986}. 
% This literature also finds significant effects of juries' racial composition on outcomes;  \cite{anwar2019jury} and \cite{hoekstra2021} find  effects of gender composition.
This evidence justifies our assumption that parties perceive different groups as having polarized distributions of conviction probabilities. 

The lack of random variation in jury selection procedures makes it difficult for the empirical literature to provide credible evidence over the effects of the choice of procedure. 
Focusing on the number of challenges, \cite{diamondAchievingDiversityJury2009} show that larger juries are more representative of the pool's demographic.\footnote{The study takes advantage of a feature of civil cases in Florida where juries are made of six jurors unless one of the parties requests a jury of twelve jurors and pays for the costs associated with such a larger jury.}
In Section \ref{sec:number}, we show that limiting the number of challenges (while keeping the number of selected jurors fixed) can have a similar effect, though at the expense of a less effective exclusion of extreme jurors.

%%%%%%%%%%%%%%%%%%%%%%%%
%%%%%%%%%%%%%%%%%%%%%%%%
%%%%%%%%%%%%%%%%%%%%%%%%
\section{Model}\label{sec:model}
%%%%%%%%%%%%%%%%%%%%%%%%
%%%%%%%%%%%%%%%%%%%%%%%%
%%%%%%%%%%%%%%%%%%%%%%%%

There are two parties to a trial, the \textbf{defendant}, $D$, and the \textbf{plaintiff}, $P$.
The outcome of the trial is decided by a jury of $j$ jurors who must be selected from the population which is composed of two groups, $a$ and $b$, in proportions $(r, 1-r)$, respectively.
The parties share a common belief about the probability that a juror $i$ will vote to convict the defendant. We denote this probability 
$c_i \in [0,1]$. Jurors of group $g\in\{a,b\}$ draw this probability independently from the same random variable $C_g$, with probability distribution $f_g(c),\ g\in\{a,b\}$. We assume that these distributions are continuous and to simplify the notation, we also assume that the boundaries of the support of $C$ are $0$ and $1$.\footnote{
This assumption is without loss of generality and all our results hold if $C$ is re-scaled in such a way that $F(c) = 0$ or $[1-F(1-c')] = 0$ for some $c,c'\in (0,1)$.} 
We denote the population distribution with $f(c)= r f_a(c) + (1-r)f_b(c)$), and the corresponding cumulative distributions with $F_g(c),\ g\in\{a,b\}$ and $F(c).$\footnote{\label{foot:uneven}Empirical evidence shows that  that parties use their challenges unevenly across groups (see the Related Literature section of the Introduction).
} 

Although throughout conviction probabilities and their distributions across groups should only be viewed as representing the parties common-\emph{beliefs}, we henceforth lighten the terminology and speak directly of conviction probabilities (rather than parties' \emph{beliefs} about conviction probabilities).

Following the literature \citep{brams_optimal_1978,flanagan_peremptory_2015}, we assume that during jury selection the parties do not account for the process of jury deliberations and, perhaps to cope with the complexity of jury selection, view the jurors' conviction probabilities as independent from one another.\footnote{See \cite{gerardi2007deliberative} and \cite{iaryczower2018can} for cases where jury deliberations have an impact on outcomes.} 
Since conviction in most U.S. trials requires a unanimous jury, the parties assume that a jury composed of jurors with conviction probabilities $\{c_i\}_{i=1}^{j}$ convict the defendant with probability $\Pi_{i=1}^j c_i$.
The defendant, therefore, aims at minimizing the product $\Pi_{i=1}^j c_i$ while the plaintiff wants to maximizing it.

To influence the composition of the jury, the defendant and the plaintiff are allowed to challenge (veto) up to $d$ and $p$ of the jurors in a \textbf{panel} of $n= j + d + p$ potential jurors randomly and independently drawn from the \textbf{population} (sometimes also called the \textbf{pool}).\footnote{%
\label{foot:venire}%
In the legal literature, what we call ``panel" is sometimes called ``\emph{venire}" (though terminology varies and the latter term is sometimes used to speak of what we call the population).
}
To avoid trivial cases, we assume throughout that $d,p \geq 1$.
The parties use these challenges in the course of a \textbf{veto procedure} $M$ (formally, an extensive game-form).
The jury resulting from the procedure is called the \textbf{effective jury}.

The two veto procedures we study are the \textbf{STRuck} procedure (\STR) and the \textbf{Strike And Replace} procedure (\SAR).
For comparison, we also consider the \textbf{Random} procedure (\RAN) which simply draws $j$ jurors independently at random from the population.
In all procedures, we assume that once a potential juror $i$ is presented to the parties, the parties observe the realized value of $c_i$ for that juror.\footnote{
    The assumption that parties have the same assessment of the probability a juror will vote for conviction is motivated by the practice of letting parties extensively question potential jurors in the \emph{voir dire} process.
    This process typically occurs in the presence of all parties, who therefore have access to the same information about the jurors' demographics, background, and opinions. 
}
The two procedures however differ in the timing with which jurors are presented to the parties.

Under \STR, the entire panel of $j+d+p$ potential jurors is presented to the parties \emph{before} they have the opportunity to use any of their challenges.
Each party, therefore, observes the value of $c_i$ for every panel member.
The defendant and the plaintiff choose to challenge up to $d$ and $p$ members of the panel, respectively. If parties do not use all their challenges, additional prospective jurors are randomly excluded from the pool until the jury size is equal to $j$.
%In practice, there are several types of \STR procedures that differ in the way the parties exercise their challenges after having questioned the jurors in the panel.
%For concreteness and tractability, we focus in this paper on the \STR procedure in which the parties have a single opportunity to exercise their challenges on the whole panel.
In equilibrium, the plaintiff challenges the $p$ jurors in the panel with lowest conviction probabilities, and the defendant challenges the $d$ jurors with highest conviction probabilities.\footnote{
Alternative methods used in the field include procedures in which the parties challenge sequentially out of subgroups of jurors from the panel only.
As long as the procedure remains of the struck type (i.e., the entire panel --- and not only the first subgroup --- is questioned before the parties start exercising their challenges), the equilibrium effective jury is often the same as under the \STR procedure we consider here.
Other outcome-irrelevant aspects of the equilibrium might, however, be different such as the number of challenges used by the parties (e.g., if the first group is made of the $j$ ``middle" jurors in the panel, they may in some cases be selected as effective jurors without the parties exercising any of their challenges).
}
Whether these challenges happen simultaneously or sequentially has no impact on the equilibrium of \STR and our results therefore apply in either case.\footnote{
Since $C$ is continuous, the probability that two jurors in a panel have the same conviction probability and one of the parties does not use all of its challenges in equilibrium has measure zero and this eventuality can therefore be neglected.
}

Under \SAR, groups of potential jurors are randomly drawn from the population and sequentially presented to the parties.
In contrast with \STR procedures, the parties must exercise their challenges on jurors from a given group \emph{without} knowing the identity of jurors from subsequent groups.
There is variation among \SAR used in practice in the size of the groups that are presented in each round.\footnote{
As well as in the ability of the parties to challenge, in a later round, potential jurors who were left unchallenged in previous rounds, a practice known as ``\emph{backstricking}".
}
For concreteness and tractability, we focus in this paper on the \SAR procedure in which jurors are presented to the parties \emph{one at a time}. 
The defendant and the plaintiff start the procedure with $d$ and $p$ challenges left, respectively.
After each draw, the plaintiff and the defendant observe the potential juror's conviction probability and, if they have at least one challenge left, choose whether or not to challenge the juror.  
If a juror is not challenged by either party, it becomes a member of the effective jury. 
Any challenged juror is dismissed and the number of challenges available to the challenging party is decreased by one. 
The process continues until an effective jury of $j$ members is formed.

The (subgame perfect) equilibrium of \SAR was characterized by \cite{brams_optimal_1978} and takes the form of threshold strategies.
In every subgame, $D$ challenges the presented juror $i$ if $c_i$ is above a certain threshold $t_D$, $P$ challenges $i$ if $c_i$ is below some threshold $t_P$, and neither of the parties challenges $i$ if $c_i \in [t_P, t_D]$.\footnote{
Each subgame can be characterized by 
the number of jurors $\kappa$ that remain to be selected, 
the number of challenges left to the defendant $\delta$, and 
the number of challenges left to the plaintiff $\pi$.
The parties threshold in subgame $(\kappa, \delta, \pi)$ are a function of the value of subgames $(\kappa -1, \delta, \pi)$, $(\kappa, \delta -1, \pi)$, and $(\kappa, \delta, \pi-1)$ (which are all possible successors to the parties action in $(\kappa, \delta, \pi)$) and the distribution of $C$, see \cite{brams_optimal_1978}.
} We will sometimes refer to these values as \textit{challenge thresholds}.
As \cite{brams_optimal_1978} show, in any subgame, $t_P < t_D$ which implies that a challenge to the same juror by both parties never occurs in equilibrium.
The equilibrium is therefore unaffected by the order in which the parties decide whether to challenge the presented juror.

In our description of \SAR, Nature moves in each round by presenting to the parties a new potential juror drawn from the population.
To facilitate comparisons between \STR and \SAR, it will sometimes be useful to consider an equivalent description of \SAR in which Nature first draws a panel of $n$ jurors $\{c_1, \dots, c_n\}$ (which the parties are not aware of) and in each round $k$ presents juror $c_k$ to the parties. 
% \footnote{
% Formally, in each round $k$ where juror $c_k$ is presented to the parties, all possible configurations of the leftover panel $\{c_{k+1}, \dots, c_n\}$ lie in the same information sets.)
% }
For similar purposes, it will sometimes be useful to view \RAN as first drawing a panel of $n$ jurors and then (uniformly at random) selecting $j$ jurors among these $n$ to form the effective jury.

%%%%%%%%%%%%%%%%%%%%%%%%
%%%%%%%%%%%%%%%%%%%%%%%%
\section{Excluding extremes and representation of minorities: An illustrative example}\label{sec:example}
%%%%%%%%%%%%%%%%%%%%%%%%
%%%%%%%%%%%%%%%%%%%%%%%%

To illustrate the differences between the two procedures, consider the simple case $d = p = j = 1$ together with distributions $C_a \sim U[0,0.5]$ and $C_b \sim U[0.5,1]$.
Also, suppose that $r = 0.1$, i.e., there is a minority of 10\% of group-$a$ jurors in the population.

Let $U_x^n[0,1]$ denote the $x$-th order statistic for a $U[0,1]$ random sample of size $n$.
With this notation and these parameters, Figure \ref{ex:STR} shows the equilibrium under \STR.
The initial node illustrates the joint distribution $C= 0.10 * C_a + 0.9 * C_b$. 
The numbers on each arrow indicate the probability of drawing a panel with the group-composition represented in the pointed boxes (conditional on each panel composition, the circled letter in the box corresponds to the group-membership of the selected juror).
Dashed arrows correspond to outcomes that lead to the selection of a group-$a$ juror 
%The text under each box indicated the selected juror's type, 
and the graph underneath each box shows the distribution of conviction probabilities of the selected juror. 

Observe that in this example, if there are group-$a$ jurors in the panel, one of them is systematically challenged by the plaintiff.
Therefore, for a group-$a$ juror (i.e., a minority juror) to be selected under \STR, there need to be at least two group-$a$ jurors in the panel of $n = 3$ presented to the parties. 
This occurs with probability about $0.03$.

\begin{figure}[t]
\centering
\caption{Illustrative example, equilibrium outcomes under \STR}
\label{ex:STR}

\resizebox{4.5in}{!}{

\begin{tikzpicture}[
    %sat/.pic = {\draw[very thin]  (0,0) plot[domain=0:.5] ({\x}, { 2*\x*(1-2*\x)*12})}
    rectangle/.style = {fill=white, draw=black, shape=rectangle,align=center, minimum width = 2.5cm, minimum height= .6cm},
    rarrow/.style={->,dashed},
    axesSt/.style={very thin, black},
    axes/.pic={
    		\draw[axesSt] (0,0) -- (2cm,0) node[right,below] {\tiny{1}} node[midway,below] {\tiny 0.5};
    		\draw[axesSt] (0,0) -- (0,1cm);},
    pics/distro/.style = {
        code={
        \begin{scope}[x=2cm,y=.4cm]
            \node[] at(-1,1) { 3 draws from $\rightarrow$};
            \node[] at (-0.15,0.2) {\tiny 0.2};
            \node[] at (-0.15,1.8) {\tiny 1.8};
            \node[] at(0.25,0.5) {\scriptsize $a$};
            \node[] at (0.75,2.1) {\scriptsize $b$};
            \draw pic {axes} (0,0);
            \draw[olive,thick] (0,0.2) -- (0.5,0.2);
            \draw[darkorange,thick] (0.5,1.8) -- (1,1.8);
        \end{scope}
        }    
        }
    ]    
    \usetikzlibrary{fit,positioning}
        \pic at (0,4) {distro};
    	\node [] (0) at (0,3.8)  {};
		\node [rectangle, fill={\oliveq}] (aaa) at (-4.5, 1) {\color{black} $a$ \textcircled{$a$} $a$};
		\node [rectangle, fill={\oliveq}] (aab) at (-1.5, 1) {{\color{black} $a$ \textcircled{$a$}} $b$};
		\node [rectangle,, fill= \darkorangeq] (abb) at (1.5, 1) {{\color{black} $a$ }\textcircled{$b$} $b$};
		\node [rectangle,, fill= \darkorangeq] (bbb) at (4.5, 1) {$b$ \textcircled{$b$} $b$};
        \node [below = .3cm of aaa] {\scriptsize $U_2^3[0,.5]$};
        \node [below = .3cm of aab] {\scriptsize $U_2^2[0,.5]$};
        \node [below = .3cm of abb] {\scriptsize $U_1^2[.5,1]$};
        \node [below = .3cm of bbb] {\scriptsize $U_2^3[.5,1]$};
         
     %   \begin{scope}[x=2cm,y=.2cm]
%            \draw pic {distro} (0,0);
	%	\end{scope}
		\draw [rarrow] (0.center) -- (aaa) node [midway,fill=white] {.001};
		\draw [rarrow] (0.center) -- (aab) node [midway,fill=white] {.027};
		\draw [->] (0.center) -- (abb) node [midway,fill=white] {.243};
		\draw [->] (0.center) -- (bbb) node [midway,fill=white] {.729};
		\begin{scope}[x=2cm,y=.2cm,shift={(-5.3cm,-.9cm)}]
            \node[gray!75] at (-0.15cm,0.6cm) {\tiny 3};
            \draw pic {axes} (0,0);
    		\draw[darkorange,thick] (1cm,0) -- (2cm,0);
            \draw[olive,thick] (0,0) plot[domain=0:.5] ({\x}, { 2*\x*(1-2*\x)*12}); 
        \end{scope}
        \begin{scope}[x=2cm,y=.2cm,shift={(-2.3cm,-.9cm)}]
            \node[gray!75] at (-0.15cm,0.8cm) {\tiny 4};
            \draw pic {axes} (0,0);
    		\draw[darkorange,thick] (1cm,0) -- (2cm,0);
            \draw[olive,thick] (0,0) plot[domain=0:.5] ({\x}, { 8*\x});
        \end{scope}
        \begin{scope}[x=2cm,y=.2cm,shift={(0.7cm,-.9cm)}]
            \node[gray!75] at (-0.15cm,0.8cm) {\tiny 4};
            \draw pic {axes} (0,0);
    		\draw[olive,thick] (0,0) -- (1cm,0);
            \draw[darkorange,thick] (0,0) plot[domain=.5:1] ({\x}, {8-8*\x});
        \end{scope}
        \begin{scope}[x=2cm,y=.2cm,shift={(3.7cm,-.9cm)}]
            \node[gray!75] at (-0.15cm,0.6cm) {\tiny 3};
            \draw pic {axes} (0,0);
    		\draw[olive,thick] (0,0) -- (1cm,0);
            \draw[darkorange,thick] (0,0) plot[domain=.5:1] ({\x}, {2*(\x-0.5)*(1-2*(\x-0.5))*12});
        \end{scope}
	
\end{tikzpicture}
}%end resizebox
%\caption*{\normalfont  \footnotesize \emph{Note:} The figure describes the equilibirum of \STR assuming  $j=p=d=1$, $C_a \sim U[0,0.5]$, $C_b \sim U[0.5,1]$, and $r = 0.10$.
%The initial node illustrates distribution $C= 0.10 * C_a + 0.9 * C_b$. 
%The numbers on each arrow indicate the probability of drawing a panel with the group-composition represented in the pointed boxes (conditional on each panel composition, the circled letter in the box corresponds to the group-membership of the selected juror).
%Dashed arrows correspond to outcomes that lead to the selection of a group-$a$ juror 
%%The text under each box indicated the selected juror's type, 
%and the graph underneath each box shows the distribution of conviction probabilities of the selected juror. 
%}
\end{figure}

\begin{figure}[t]
\caption{Illustrative example, equilibrium strategies and outcomes under \SAR} \label{fig:SAR}
\centering
\resizebox{4.5in}{!}{
\begin{tikzpicture}[
    axesSt/.style={very thin, black},
    axes/.pic={
    		\draw[axesSt] (0,0) -- (2cm,0) node[right,below] {\tiny{1}} node[midway,below] {\tiny 0.5};
    		\draw[axesSt] (0,0) -- (0,1cm);},
    pics/distro/.style = {
        code={
        \begin{scope}[x=2.2cm,y=.44cm]
            \node[] at(0.5,3.5) { $\underset{\downarrow}{\text{Each round 1 draw from}}$};
            \node[] at (-0.15,0.2) {\tiny 0.2};
            \node[] at (-0.15,1.8) {\tiny 1.8};
            \node[] at(0.25,0.5) {\scriptsize $a$};
            \node[] at (0.75,2.1) {\scriptsize $b$};
            \draw pic {axes} (0,0);
            \draw[olive,thick] (0,0.2) -- (0.5,0.2);
            \draw[darkorange,thick] (0.5,1.8) -- (1,1.8); 
        \end{scope}
        }    
        },
    rectangle/.style={fill=white, draw=black, shape=rectangle,align=center, minimum width = 2.5cm},
    circle/.style={fill=white, draw=black, shape=circle},
    ]
    
    \pic at (15,18) {distro};

	\node [circle] (0) at (21, 19) {Round 1};
	\node [rectangle] (1) at (26.5, 16.0) {$c_i\in[.788,1]$ \\ D challenges};
	\node [rectangle, fill={\darkorangeq}, label={below: \textbf{\color{darkorange} Group-$b$}}] (2) at (21, 16.0) { $c_i\in[.619,.788]$ \\ No challenges  };
	\node [rectangle] (3) at (15.6, 16.0) { $c_i \in [0, .619]$ \\ P challenges};
	\node [circle] (12) at (21, 13.00) {Round 2};
	\node [rectangle] (13) at (18.3, 13.00) { $c_i \in [.70, 1]$ \\ D challenges };
	\node [rectangle, fill={\oliveq},  label={[align=center] below: \textbf{\color{olive} Group-$a$}}]  (cc) at (12.8, 13.00)
	        { $c_i\in[0, .50]$  \\ No challenges };
	\node [rectangle, fill={\darkorangeq}, label={[align=center] below: \textbf{\color{darkorange} Group-$b$}}] (14) at (15.6, 13.00) { $c_i\in[.5, .70]$  \\ No challenges };
	\node [rectangle,fill={\darkorangeq},  label={below: \textbf{\color{darkorange} Group-$b$}}] (16) at (26.5, 13.00) { $c_i \in [.70, 1]$ \\\ No challenges };
	\node [rectangle] (17) at (23.7, 13.00) { $c_i \in [0, .70]$ \\ P challenges };
	\node [rectangle,fill={\darkorangeq},  label={below: \textbf{\color{darkorange} Group-$b$}}] (22) at (26.5, 10.00) { $c_i\in[.5,1]$ };
	\node [rectangle, fill={\darkorangeq}, fill={\oliveq}, label={below: \textbf{\color{olive} Group-$a$}}] (24) at (23.7, 10.00) { $c_i\in[0,.5]$ };
	\node [rectangle, fill= \darkorangeq, label={below: \textbf{\color{darkorange} Group-$b$}}] (26) at (18.3, 10.00) { $c_i\in[.5,1]$ };
	\node [rectangle, fill={\oliveq}, label={below: \textbf{\color{olive} Group-$a$}}] (28) at (15.6, 10.00) { $c_i\in[0,.5]$ };
	\node [style=circle] (30) at (21, 10.00) {Round 3};

		\draw [->, dashed]     (0) -- (1) node [midway,fill=white] {.3816};
		\draw [->, in=90, out=-90]  (0) -- (2) node [midway,fill=white] {.3042};
		\draw [->, dashed]     (0) -- (3) node [midway,fill=white] {.3142};
		\draw [->]             (1) -- (16) node [midway,fill=white] {.54};
		\draw [->, dashed]     (1) -- (17) node [midway,fill=white] {.46};
		\draw [->]             (3) -- (13) node [midway,fill=white] {.54};
		\draw [->]             (3) -- (14) node [midway,fill=white] {.36};
		\draw [->,dashed]      (3) -- (cc) node [midway,fill=white] {.10};
		\draw [->]             (13) -- (26) node [midway,fill=white] {.90};
		\draw [->, dashed]     (13) -- (28) node [midway,fill=white] {.10};
		\draw [->]             (17) -- (22) node [midway,fill=white] {.90};
		\draw [->, dashed]     (17) -- (24) node [midway,fill=white] {.10};

\end{tikzpicture}
} % end resizebox
\smallskip

%\caption*{\normalfont \footnotesize \emph{Note:} The figure describes the equilibrium strategies conditional on the conviction probability of the juror drawn in each round for the case $j=d=p=1$, $C_a \sim U[0,0.5]$, $C_b \sim U[0.5,1]$ and $r = 0.10$. 
%Dashed arrows correspond to paths that may lead to the selection of a group-$a$ juror. 
%The numbers on each arrow indicate the probability of the path conditional on reaching the previous node. 
%The second row of text inside the boxes indicates an equilibrium action (in round 3 challenges are exhausted and the parties do not take any action); bold text below the boxes indicates the group of the selected juror in the game outcome.}
\end{figure}

In contrast, a group-$a$ juror can be selected under \SAR even if the panel contains a single group-$a$ juror.
To understand why, consider the equilibrium of \SAR which is illustrated in Figure \ref{fig:SAR}.
In the figure,  dashed arrows correspond to paths that may lead to the selection of a group-$a$ juror, and the numbers on each arrow indicate the probability of the path conditional on reaching the previous node. 
The second row of text inside the boxes indicates an equilibrium action (in round 3 challenges are exhausted and the parties do not take any action); bold text below the boxes indicates the group of the selected juror in the game outcome.

If a group-$b$ prospective juror with a sufficiently low conviction probability (that is, $c_i \in [0,0.619]$) is presented first, then it will be challenged by the plaintiff.
This leads to a subgame in which only the defendant has challenges left and a group-$a$ juror is more likely to be selected than if a juror was randomly drawn from the population.
In particular, any group-$a$ juror presented at the beginning of this later subgame is left unchallenged by the defendant and selected to be the effective juror (even if this juror is the only group-$a$ juror in the panel because the third juror --- who, in this case, is never presented to the parties --- happens to be a group-$b$ juror).
This course of action follows from $P$'s choice to challenge a group-$b$ juror with low conviction probability in the first round, which leaves $P$ without challenges left in the second round.
This choice of $P$ is optimal from the perspective of the first round of \SAR (\emph{before} the plaintiff learns that the second juror in the panel is a group-$a$ juror), but suboptimal under \STR where, having observed the conviction probability of all jurors in the panel, the plaintiff would have challenged the group-$a$ juror instead.

Considering only the branch of the \SAR game-tree that starts with a challenge from $P$, the probability of selecting a group-$a$ juror is  $ 0.31 * (0.54 * 0.1 + 0.10)$ $\approx 0.05$
Adding the possibility that a minority juror is selected after $D$ challenges in the first round followed by a challenge from $P$ in the second round (which occurs with probability $0.38 * 0.46 * 0.1 \approx 0.02$), the probability of selecting a minority juror under \SAR is 0.066.\footnote{%
These are the only cases in which a minority juror can be selected under \SAR.
In particular, jurors accepted in the first round are always group-$b$ jurors ($c_i \in [0.619, 0.788]$). 
So are jurors accepted in the second round following a challenge from $D$ is the first round ($c_i \in [0.70, 1]$).} 
This is larger than the probability under \STR, 0.03, yet smaller than under $RAN$, 0.10.

\smallskip

In this example, the better representation of minority jurors produced by \SAR comes at the expense of selecting more extreme jurors.
Suppose for the sake of illustration that jurors are considered extreme if they come from the top or bottom 5th percentile of $C$.
In our example, the bottom and top 5th percentile correspond to conviction probabilities below 0.25 and above 0.94, respectively.
The selected juror is within the bottom range with probability 
$0.015$ under \STR versus
$0.033$ under \SAR,
and in the top range with probability
$0.076$ under \STR  versus
$0.083$ under \SAR.

To understand the source of these differences, consider the bottom 5th percentile $[0,0.25]$ (a symmetric explanation applies to the \emph{top} 5th percentile).
As indicated in Figure \ref{ex:STR}, when \STR selects a group-$a$ juror --- the type of juror whose conviction probability could possibly be in the bottom 5th percentile --- the distribution of that juror's conviction probability follows the middle or upper order-statistics of a random sample from $C_a$.
These order-statistics are unlikely to result in the selection of a juror with conviction probability in the bottom 5th percentile.
In contrast, as Figure \ref{fig:SAR} illustrates, 
all paths leading \SAR to select a group-$a$ juror result in the juror's conviction probability being drawn from $U[0,0.5]$ itself,
which makes \SAR more likely to select a juror in the bottom 5th percentile than \STR.

In the next two sections, we investigate the extent to which the advantages of \SAR in terms of minority-representation and of \STR in terms of exclusion of extreme generalizes beyond this illustrative example.

%%%%%%%%%%%%%%%%%%%%%%%%
%%%%%%%%%%%%%%%%%%%%%%%%
%%%%%%%%%%%%%%%%%%%%%%%%
\section{Exclusion of extremes}\label{sec:extreme}
%%%%%%%%%%%%%%%%%%%%%%%%
%%%%%%%%%%%%%%%%%%%%%%%%
%%%%%%%%%%%%%%%%%%%%%%%%

The peremptory challenge procedures implemented in U.S. jurisdictions are often viewed as a way to foster impartiality by preventing extreme potential jurors from serving on the effective jury.\footnote{
	 See Footnote \ref{foot:partial} and its associated quote.
	 For legal arguments in favor of peremptory challenges based on the Sixth Amendment, see, among others, \cite{beckCurrentStatePeremptory1998}, \cite{biedenbenderHollandIllinoisSixth1991}, \cite{bonebrakeSixthFourteenthAmendments1988}, \cite{horwitz_extinction_1992-1}, and \cite{keene_fairness_2009}.
	 }
In the context of our model, we interpret this goal as that of limiting the presence in the jury of jurors from the tails of the distributions of conviction probabilities.

We define a juror $i$ as \emph{extreme} if its conviction probability $c_i$ lies below or above given thresholds (see Section \ref{sec:ext} for results under an alternative definition).
For brevity, we will focus on jurors who qualify as extreme because their conviction probability lies \emph{below} some threshold $\ul{c} > 0$.
All our results about extreme jurors apply symmetrically to jurors whose conviction probability lies \emph{above} a given threshold $\ol{c} < 1$.

In the previous section's example, jurors in the bottom 5th percentile of $C$ are selected less often under \STR than \SAR.
% This is not true in general.
% Fixing a particular threshold $\ul{c} > 0$ --- or percentile of $C$ --- to characterize jurors as extreme, there always exists distributions of $C$ and values of $d$, $p$, and $j$ such that \SAR selects fewer extreme jurors than \STR.
Our first result formally characterizes the conditions under which this outcome holds, showing there always exists a sufficiently small threshold such that the probability of selecting  
extreme jurors (i.e., below that threshold) is greater under \SAR than under \STR. 
\begin{table}[t]
    \caption{Notation reference}
    \centering
    \renewcommand{\arraystretch}{1.4}
    \begin{tabular}{|ll|ll|}
    \hline
         $j$ &jury size               &   $\ul{\T}_{M}(x ; c)$ & prob. at least $x$ jurors with $c_i<c$ \\
         $d+p$ &peremptory challenges &   ${N}^c_M$             &expected n. of jurors with $c_i<c$\\
         $n=j+d+p$ &panel size        &   $\M_M(x)$           &prob. at least $x$  minority in jury \\ 
         $c$&   conviction probability &\multicolumn{2}{|l|}{procedure \small{  $M\in \{\STR, \SAR, \RAN \}$}} \\ \hline      
    \end{tabular}
\end{table}

Let $\ul{\T}_M(x ; c)$ denote the probability that there are at least $x$ jurors with conviction probability \emph{smaller or equal} to $c$ in the jury selected by procedure $M$.
% Conversely, let $\ol{\T}_M(x ; c)$ denote the probability that there at least $x$ jurors with conviction probability \emph{larger or equal} $c$ in the jury selected by procedure $M$.

\begin{prop}
	\label{prop:n_extreme}
	For any $x\in \{1, \dots, j\}$,
	there exists $\ul{c} > 0$ 
	such that 
 	$\ul{\T}_{STR}(x ; c) < \ul{\T}_{\SAR}(x ; c)$  for all $c \in (0, \ul{c})$.
%  	\footnote{
%  	To illustrate the symmetry that applies to all our results about extreme jurors, let $\ol{\T}_M(x ; c)$ denote the probability that there are at least $x$ jurors with conviction probability \emph{larger or equal} to $c$ in the jury selected by procedure $M$.
%  	Then, it is also the case that for any $x\in \{1, \dots, j\}$,
% 	there exists $\ol{c} < 1$ 
% 	such that 
%  	$\ol{\T}_{STR}(x ; c) < \ol{\T}_{\SAR}(x ; c)$  for all $c \in (\ol{c}, 1)$.}
% 	LINK TO PROOF: \ref{app:n_extreme}
	\end{prop}

All proofs are in the appendix. 
A symmetric statement, which we omit, applies to extreme jurors at the right-end of the distribution.
Note that Proposition \ref{prop:n_extreme} can be rephrased in terms of stochastic dominance.
Let ${N}^c_M$ denote the expected number of jurors of type $c_i \leq c$ in the jury selected by procedure $M$.
Then, Proposition \ref{prop:n_extreme} says that there exists $\ul{c} > 0$ such that ${N}^c_{\SAR}$ has first-order stochastic dominance over ${N}^c_{STR}$  for all $c \in (0, \ul{c})$.
A direct corollary of Proposition \ref{prop:n_extreme} is therefore that the \emph{expected number} of extreme jurors is larger under \SAR than under \STR.

For an intuition about Proposition \ref{prop:n_extreme}, consider the case $x = 1$.
As illustrated in Section \ref{sec:example}, the panel must be composed of more than one extreme juror for \STR to select at least one such juror (since, if there is a single extreme juror in the panel, that juror is systematically challenged by the plaintiff).
In contrast, even in panels with a single extreme juror, the extreme juror can be part of the effective jury resulting from \SAR.
This happens, for example, if the extreme juror is presented to the parties after they have both exhausted all their challenges. 
The single extreme juror can also be accepted by both parties if its conviction probability is sufficiently close to $\ul{c}$ and it is presented after the plaintiff used most of its challenges on non-extreme potential jurors.\footnote{
Subgames in which the defendant has more challenges left than the plaintiff can lead the plaintiff to be conservative and accept jurors who are ``barely extreme" $(c_i \approx \ul{c})$ in order to save its few challenges left for ``very extreme" jurors  ($c_i \approx 0$).
}
The proof then follows from the fact that, as $\ul{c}$ tends to zero, the probability that the panel contains more than one extreme juror goes to zero faster than the probability the panel contains a single extreme juror.\footnote{
Proposition \ref{prop:n_extreme} crucially depends on averaging across all possible panels and does \emph{not} state that \STR rejects more extreme jurors than \SAR for \emph{any} particular realization of the panel.
The latter would obviously imply Proposition \ref{prop:n_extreme} but turns out to be false in general. 
For a counterexample, let $j=d=p=1$. 
Consider a panel of three jurors with $c_{2} < c_{3}<\ul{c}$ and $\ul{c} < c_{1} < \ol{c}$ and where the index of the jurors indicate the order in which they are presented under \SAR.
For this panel, \STR always leads to the selection of extreme
juror $3$. 
In contrast, provided $c_{2}$ falls between the challenge thresholds of the defendant and the plaintiff in the first round (which happens with positive probability), \SAR selects non-extreme juror $1$.
}

\begin{figure}[t]
    \centering
    \caption{Distributions of conviction probabilities by group under extreme, moderate, and mild group-polarization}
    \label{fig:betaPDFs}
    \includegraphics[width=4.5in]{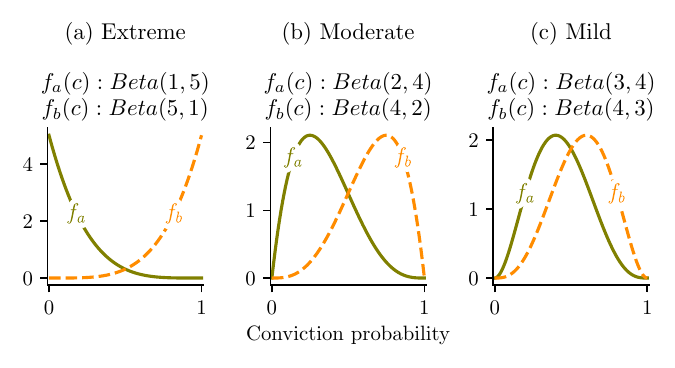}
\end{figure}

Proposition \ref{prop:n_extreme} is silent about the value of the threshold $\ul{c}$ below which \STR selects fewer jurors than \SAR, as well as the size of $\ul{\T}_{\SAR}(x ; c) - \ul{\T}_{STR}(x ; c)$ for $c< \ul{c}$. 
These values depend on the model's parameters.
To illustrate that the advantage of \STR over \SAR can be large, even with relatively high thresholds defining extremes, we simulate $\ul{\T}_{\STR}(1 ; c)$  and $\ul{\T}_{\SAR}(1 ; c)$ 
using $j=12$, $d=6$, and $p=6$, a typical combination of jury size and number of peremptory challenges in U.S. jurisdictions. 
For the distribution of conviction probabilities in the population, we use mixtures of beta distributions that represent a population made of two groups with polarized views, with $r=0.25$ unless otherwise noted.
Although the results in this section do not depend on whether jurors come from  polarized groups, using these distributions facilitates comparisons with Section \ref{sec:grouprep} where we study group-representation. 
The distributions we adopt are illustrated in Figure \ref{fig:betaPDFs}, which are meant to represent extreme, moderate, and mild levels of polarization (Panels (a), (b), and (c) respectively).
Additional simulations results using $U[0,1]$ are reported in the online appendix (\cite{mvdl-external}).

\begin{figure}[t]
    \centering
    \caption{Fraction of juries with at least one extreme juror}
    \label{fig:atleast1-3betas}
	\includegraphics[width=4.5in]{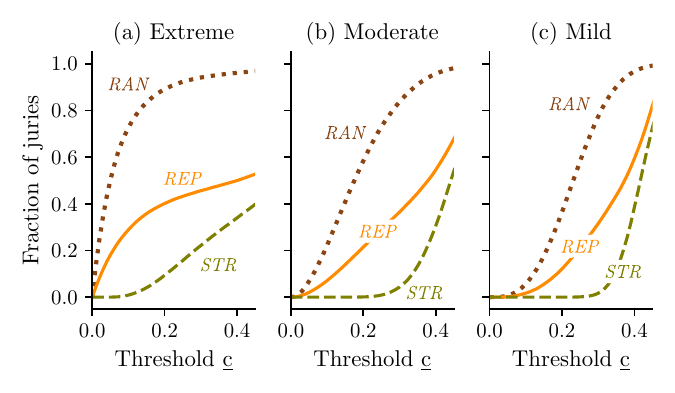} 
	%  \caption*{\footnotesize \normalfont \emph{Note:} For each set of parameters, results on the vertical axis are averages across 50,000 simulated jury selections, fixing $j=12$, $d= p=6$, and $C \sim  0.25 * C_a + 0.75 * C_b$ throughout (distributions $C_a$ and $C_b$ illustrated in Figure \ref{fig:betaPDFs}). 
	%  Each line illustrates the fraction of juries with at least one extreme juror, where a juror is considered extreme if her conviction probability falls below the threshold \ul{c} corresponding to the value on the horizontal axis.
	 %}
\end{figure}

In simulations using these parameters, we found \STR to exclude more extreme jurors than \SAR even when the threshold for defining jurors as extreme is relatively high. In Figure \ref{fig:atleast1-3betas} each line illustrates the fraction of simulated juries with at least one extreme juror, where a juror is considered extreme if her conviction probability falls below the threshold \ul{c} corresponding to the value on the horizontal axes. Results in this and all the folllowing simulations are averages across 50,000 simulated jury selections.

The difference between the propensity of \STR and \SAR to select extreme jurors  is sizable. 
For example, in all three sets of simulations, less than 1\% of juries selected by \STR include at least one juror with conviction probability below the 10th percentile of the distribution (the 10th percentile corresponds to 0.10 under the extreme polarization distribution, 0.25 under moderate polarization, and 0.28 under mild polarization).
Under \SAR, the proportion of juries with at least one juror below the 10th percentile rises to 29\% with extreme polarization, 28\% with moderate polarization, and remains quite high at 27\% even under mild polarization. 
For comparison, a random selection would have resulted in over 70\% of the juries featuring at least one such juror in all scenarios.

In these simulations, both procedures select fewer extreme jurors than a random draw from the population.
Somewhat surprisingly, this is not true in general.
There exist distributions and values of the parameters $d$, $p$, and $j$ for which \SAR selects \emph{more} extreme jurors than \RAN, no matter how small the threshold below which a juror is considered as extreme. 
In contrast, as we show in the next proposition, \STR always selects fewer extreme jurors than  \RAN.

\begin{prop}
\label{prop:extremeRAN}
    % (a) 
    For any $x \in \{0,\dots, j\}$,
	there exists $\ul{c} > 0$ such that $\ul{\T}_{STR}(x ; c) < \ul{\T}_{RAN}(x ; c)$  for all  $c \in (0, \ul{c})$.\footnote{
% 	Part (a) of 
	Proposition \ref{prop:extremeRAN} generalizes Theorem 2 in \cite{flanagan_peremptory_2015} which shows that the statement holds for $x=j$.
}
% 	In contrast, 
% 	(b) there exists distributions $f(c)$ and values of $x, j,d$, and $p$ (including values with $d = p$) such that, for some $\ul{c} > 0$, we have $\ul{\T}_{\SAR}(x ; c) >  \ul{\T}_{RAN}(x ; c)$  for all  $c \in 0, \ul{c})$.
\end{prop}

\begin{figure}[t]
    \centering
    \caption{Fraction of juries with at least one extreme juror (case in which \SAR is more likely to pick extreme jurors than \RAN)}
    \label{fig:prop2-uni}
    \includegraphics[width=3.5in]{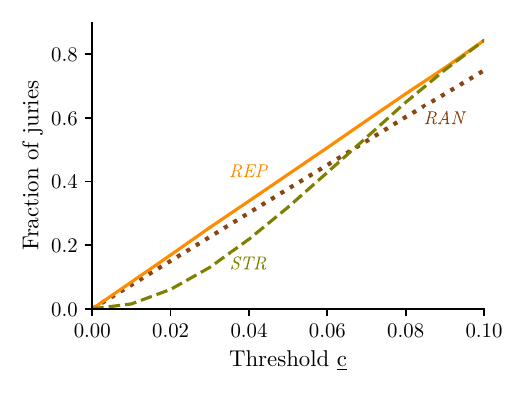}
%    \caption*{\footnotesize \normalfont \emph{Note:} 
%    For each set of parameters, results on the vertical axis are averages across 50,000 simulated jury selections, fixing $j= d = p =1$, and $C \sim 0.75 * U[0,0.1] + 0.25 * U[0.9,1]$ throughout. 
%    Each line illustrates the fraction of juries with at least one extreme juror, where a juror is considered extreme if her conviction probability falls below the threshold \ul{c} corresponding to the value on the horizontal axis.
 %   }
\end{figure}

Figure \ref{fig:prop2-uni} illustrates
Proposition \ref{prop:extremeRAN} and the fact that a similar statement does not hold for \SAR.
For the simulations in the figure, we let $j=d=p=1$ and adopt an extremely polarized distribution of conviction probabilities with $C \sim 0.75 * U[0,0.1] + 0.25 * U[0.9,1]$.
In this case (as in others), \STR excludes extreme jurors more often than \RAN because, for any realization of the panel, the juror with the lowest conviction probability is never selected under \STR (whereas the same juror is selected with positive probability under \RAN). 
Under \SAR, however, if the distribution is sufficiently right-skewed, the plaintiff is more likely than the defendant to challenge in the first round.
A challenge by the plaintiff in the first round leads to a subgame in which only the defendant has challenges left and the selection of an extreme juror is more likely than under a random draw.
When they are sufficiently large
(i) the added probability of selecting an extreme juror when the defendant has more challenges left than the plaintiff, coupled with 
(ii) the probability of a challenge by the plaintiff in the first round can, as in the simulation depicted in Figure \ref{fig:prop2-uni}, lead to \SAR selecting more extreme jurors than \RAN.

We could not fully characterize the situations in which \SAR selects more extreme jurors than \RAN, and we never observed such a situation in simulations where $C$ is a mixture of beta or uniform distributions.
The example in Figure \ref{fig:prop2-uni} (as well as other examples we found) requires extreme skewness in the distribution, which may be viewed as unlikely. 
In this sense, situations in which \SAR selects \emph{more} extreme jurors than \RAN might represent worst-case scenarios for \SAR's ineffectiveness at excluding extreme jurors.

%%%%%%%%%%%%%%%%%%%%%%%%
%%%%%%%%%%%%%%%%%%%%%%%%
\section{Representation of minorities\label{sec:grouprep}}
%%%%%%%%%%%%%%%%%%%%%%%%
%%%%%%%%%%%%%%%%%%%%%%%%

In this section, we study the extent to which  \STR's tendency to exclude more extreme jurors than \SAR impacts the representation of minorities.
Without loss of generality, we let group-$a$ be the minority.
Since the parties do not care intrinsically about group-membership, any asymmetry in the use of their challenges arises from heterogeneity in preferences for conviction between groups.
In our simulations, we assume that group-$a$ is biased in favor of acquittal in the sense that $C_b$ first-order stochastically dominates $C_a$.\footnote{We also simulated the scenario in which the minority is biased towards conviction, the results, which we report in the Appendix, are symmetrically very close.}

As suggested by Proposition \ref{prop:n_extreme}, which procedure better represents minorities strongly depends on the polarization between the two groups, and the concentration of minority jurors at the tails of the distribution of conviction probabilities.
% Without further restriction on $C_a$, $C_b$, and $r$, it is easy to find cases in which $\M_{STR}(x ; r) > \M_{SAR}(x ; r)$ as well as cases in which $\M_{STR}(x ; r) > \M_{SAR}(x ; r)$.
% To understand why, observe that as long as two procedures do not result in the same distribution of conviction probabilities, making $M_1$ select more minorities than $M_2$ is just a matter of splitting the distribution of $C$ into $C_a$ and $C_b$ in such a way that $C_a$ is concentrated around the conviction types tat $M_1$ select more often than $M_2$.

\begin{figure}[t]
    \centering
    \caption{Jury selection and minority representation in size-1 juries}
	\label{fig:counter}
	\includegraphics[width=4.5in]{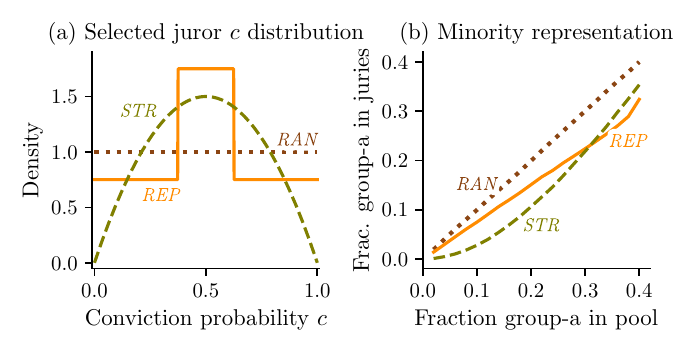}

%    \caption*{\footnotesize \normalfont \emph{Note:} 
%    For each set of parameter, results on the vertical axes are averages across 50,000 simulated jury selections, fixing $j=1$, $d= p =1$, and $C \sim r * U[0,r] + (1-r) * U[r,1]$ throughout.
%    The distribution in panel (a) is independent of $r$; the lines in panel (b) interpolate results from 20 values of $r$.}
\end{figure}

To illustrate, suppose that $d = p = j = 1$ and  $C \sim U[0,1]$.
For this case, the distributions of conviction probabilities for the juror selected under \RAN, \STR, and \SAR are displayed in Figure \ref{fig:counter}(a).\footnote{The distribution in panel (a) is independent of $r$; the lines in panel (b) interpolate results from 20 values of $r$ between 0.02 and 0.4.}
Consistent with Proposition \ref{prop:n_extreme}, below some threshold $\ul{c} \approx 0.15$, the probability of selecting a juror $i$ with $c_i < \ul{c}$ is lower under \STR than under \SAR.
If the two groups are polarized and the distribution of $C_a$ is sufficiently concentrated below $\ul{c}$, it follows that \STR selects a minority juror less often than \SAR.
But the same is not true if the distributions lack polarization or the minority is too large.
For example, let $C_a \sim U[0,r]$ and $C_b \sim U[r,1]$ so that $C \sim U[0,1] =  r U[0,r] + (1-r)U[r,1]$.
Since the parties only care about a juror's conviction probability and not about its group-membership \emph{per se}, the value of $r$ does not affect the distributions of conviction probabilities for the juror selected under \RAN, \STR, or \SAR.
However, as illustrated in Figure \ref{fig:counter}(b), low values of $r$ --- which concentrate minorities at the bottom of the distribution --- make \SAR select more minorities than \STR, whereas higher values of $r$ --- which spread the minority over a larger range of conviction-types --- make \STR select more minorities than \SAR.

From this example, we see that non-overlapping group-distributions are not sufficient to guarantee that \SAR selects more minority jurors than \STR.
Neither is making the minority arbitrarily small. 
For example, regardless of the size of the minority $r$, concentrating the support of the minority distribution inside the interval $[0.2,0.3]$ would result in \STR selecting more minorities, as can be seen from Panel (a). 
% In this case, however, it would be more difficult to claim that the minority favors the defendant, as there would be a mass of majority in the juror pool with lower conviction probabilities than all minority agents. 
However, combining a small minority with group-distributions that minimally overlap concentrates the distribution of group-$a$ at the tails which, as implied by Proposition \ref{prop:n_extreme}, makes \SAR select more minorities than \STR.

Formally, consider a sequence of triples $\{(C^i_a, C^i_b, r^i)\}_{i = 1}^\infty$.
If,
\begin{enumerate}[(i)]
	\item $r^i \in (0,1]$ for all $i \in \mathbb{N}$ with $\lim_{i \ra \infty} r^i = 0$, and
	\item $C^i_a$ and $C^i_b$ converge in distribution to $C^*_a$ and $C^*_b$, with either $\p(C^*_a < C^*_b) = 0$ or $\p(C^*_a > C^*_b) = 0$,
\end{enumerate}
then we say that \textbf{there is a vanishing minority and group-distributions that do not overlap in the limit}.
For any such sequence, let $\M^i_M(x)$ denote the probability that there are at least $x$ minority jurors in the jury selected by procedure $M$ when group-distributions are $C^i_a$ and $C^i_b$ and the proportion of minority jurors in the population is $r^i$.

\begin{prop}
	\label{prop:minor_compare}
	% (a) 
	Suppose that, under $\{(C^i_a, C^i_b, r^i)\}_{i = 1}^\infty$, there is a vanishing minority and group distributions that do not overlap in the limit.
	Then for all $x \in \{1, \dots, j\}$, there exists $k$ sufficiently large such that  $\M^i_{\SAR}(x) > \M^i_{\STR}(x)$ for all  $i > k$.\footnote{
	Note that, despite the argument presented in the motivating example illustrated in Figure \ref{fig:counter},
	Proposition \ref{prop:minor_compare} does not follow directly from Proposition \ref{prop:n_extreme}.
	The reason is that, unlike in the motivating example, most of the sequences $\{(C^i_a, C^i_b, r^i)\}_{i = 1}^\infty$ covered by Proposition \ref{prop:minor_compare} are such that $C^i = r^i C^i_a + (1-r^i) C^i_b$ varies across the sequence (i.e., $C^h \neq C^k$ for most $h,k \in \mathbb{N}$). 
	}
\end{prop}

Given the result in Proposition \ref{prop:minor_compare}, it is natural to wonder how small the minority and the overlap between the group-distributions must be for \SAR to select more minority jurors than \STR.
When the latter is true, one may also wonder about the size of $\M_{\SAR}(x ; r) - \M_{\STR}(x; r)$ is. 
Again, the answer depends on the model's parameters.
To inform these questions, we ran a set of simulations with $d=p=6$ and $j=12$ using the distributions displayed in Figure \ref{fig:betaPDFs}.

\begin{figure}[t]
    \caption{Representation of Group-a when Group-a is a minority}\label{tab:betas-grouprep}
    \centering
	\includegraphics[width=4.5in]{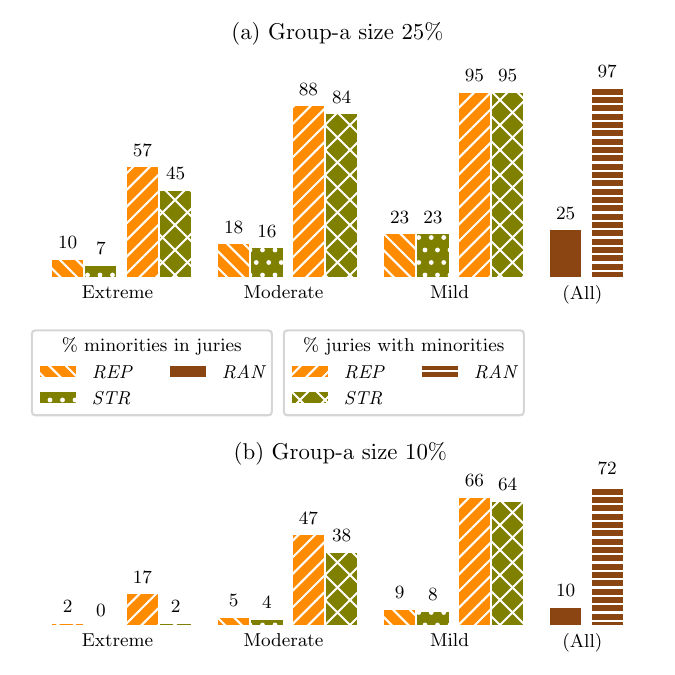}
\end{figure}

The results of our simulations are displayed in Figure \ref{tab:betas-grouprep}, reporting the percent of selected minorities in all juries, and the precent of juries with at least one minority juror. The results 
suggest that \SAR might select more minority jurors than \STR even when the size of the minority is relatively high (Panel (a) reports results with $r=.25$, Panel (b) with $r=.10$) and the overlap between the group-distributions significant.
However, without stark polarization across groups,\footnote{
Recall that $C_a$ and $C_b$ represent the parties' \emph{beliefs} that randomly drawn group-$a$ or group-$b$ jurors eventually vote to convict the defendant.
Polarized $C_a$ and $C_b$, therefore, corresponds to groups that are \emph{perceived} by the parties to have different probabilities of voting for conviction (whether or not this materializes when jurors actually vote on conviction at the end of the trial).} 
differences between the procedures' propensities to select minority jurors appear to be small.
For example, under the distributions we labeled as ``extreme group polarization'' and with minorities representing 10\% of the population, only 2.3\% of juries selected by \STR include at least one minority juror whereas this number rises to 17.1\% under \SAR (random selection would generate over 70\% of such juries). 
However, under the distributions we labeled as ``mild group polarization'', the same numbers  
become 66.5\% under \SAR and 64.5\% under \STR.

%%%%%%%%%%%%%%%%%%%%%%%%%%%%%%%%%
%%%%%%%%%%%%%%%%%%%%%%%%%%%%%%%%%
\section{Changing the number of challenges}\label{sec:number}
%%%%%%%%%%%%%%%%%%%%%%%%%%%%%%%%%
%%%%%%%%%%%%%%%%%%%%%%%%%%%%%%%%%

%So far, we have compared \STR and \SAR assuming that the number of challenges the parties can use, $d$ and $p$, was the same under each procedure. Judges often have discretion in selecting the procedure through which the parties use their challenges (see Footnote \ref{foot:flexible}), but the number of challenges that the parties can use are typically specified more rigidly by state rules of criminal procedure. 
%% REMOVED -- from what I have seen from evans data, judges can grant additional challenges if they like to do so

The number of challenges that the parties can use are typically specified by state rules of criminal procedure. 
In the last decades, several states have reduced the number of challenges the parties can use.\footnote{
For example, California's Bill 843 (2016) reduced the number of challenges a criminal defendant is entitled to from 10 to 6 (for charges carrying a maximal punishment of one year in prison, or less).
}
In some instances, these reforms also clarify or alter the jury selection procedures used in the state.\footnote{
Examples include the 2003 reform of jury selection in Tennessee where some aspects of the jury selection procedure were codified to apply uniformly across the state, while the number of peremptory challenges was also slightly reduced \citep[see][]{cohen_jury_2003}. 
} 
In the context of such broader reforms, it is natural to ask how the ability to change \emph{both} the number of challenges the parties are entitled to \emph{and} the procedure through which the parties exert their challenges affect the trade-off between the exclusion of extreme jurors and the representation of minorities.

Throughout this section, we fix an arbitrary value of $j$ and consider varying $d = p$.
For any procedure $M$, let $M\ph y$ denote the version of $M$ when $d = p = y$.
The notation for the two previous sections then carries over, with $\ul{\T}_{M\ph y}(x ; c)$ denoting the probability that at least $x$ jurors with conviction probability below $c$ are selected under $M\ph y$, and ${\M}_{M\ph y}(x)$ the probability that at least $x$ minority jurors are selected under $M\ph y$.\footnote{
Again, in the case of extreme jurors, we focus on jurors who qualify as extreme because their conviction probability falls \emph{below} a certain threshold $\ul{c}$, though all of our results hold symmetrically for jurors who qualify as extreme because their conviction probability lies \emph{above} a certain threshold $\ol{c}$,
}

\begin{figure}[t]
    \centering
    \caption{The effect of varying the number of challenges}
    \label{fig:nchallenges}
	\includegraphics[width = 4.5in]{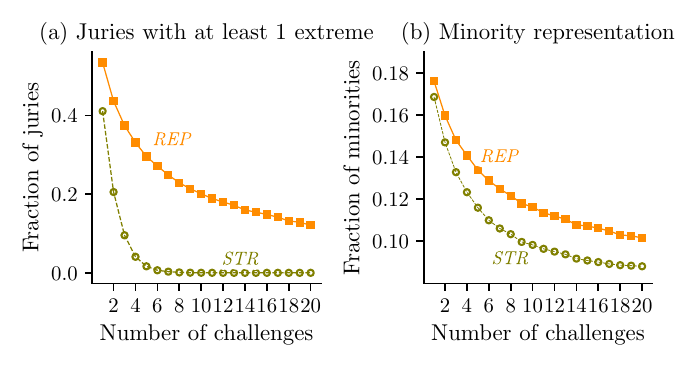}
	%    \caption*{\footnotesize \normalfont \emph{Note:} Fraction of juries with at least one juror below the 10th percentile (left panel) and fraction of minority jurors (right panel) under \STR (green starred markers) and \SAR (orange square markers). 
	%    For each set of parameters, results on the vertical axes are averages across 50,000 simulated jury selections, fixing $j=12$ and $C \sim  0.2 * C_a + 0.8 * C_b$ throughout (distributions $C_a \sim Beta(2,4)$ and $C_b \sim Beta (4,2)$, see Figure \ref{fig:betaPDFs}(b)).
	%    The values of $d=p$ are on the horizontal axes.}
		
\end{figure}

For illustration, we first present in Figure \ref{fig:nchallenges} simulations for the case with $j=12$, with $d=p$ from 1 to 20 (in the horizontal axes), and distribution $C \sim  0.2 * C_a + 0.8 * C_b$, with $C_a \sim Beta(2,4)$ and $C_b \sim Beta(4,2)$ ($C_a$ and $C_b$ are illustrated in the Figure \ref{fig:betaPDFs}(b)).
In Panel (a) we consider a juror as extreme if its conviction probability falls in the bottom 10th percentile of $C$ ($0.27$).
Unsurprisingly, the fraction of juries with at least one \emph{extreme} jurors \emph{decreases} as the number of challenges awarded to the parties increases, regardless of the procedure that is used. 
Conversely, the fraction of \emph{minority} jurors \emph{decreases} with the number of challenges under both procedures (Panel (b)).  
For both \STR and \SAR, more challenges lead to fewer extreme jurors being selected at the expense of a lower minority representation.

%As Figure \ref{fig:nchallenges}(a) illustrates, however, increasing the number of challenges decreases the selection of extreme jurors much faster under \STR (circle markers) than under \SAR (square markers). 
% This is why the trade-off between objectives we described off does not generalize if we compare across selection rules: procedure $\STR\ph w$ could both select more minority and fewer extreme jurors than $\SAR\ph y$, which is implied by the following proposition:
As a consequence, for all values of $y \in \{2,\dots, 18\}$, there exists $w < y$ such that $\STR\ph w$ performs better than $\SAR\ph y$ in terms of \emph{both} objectives.\footnote{
Specifically, in this example, for any $y \in \{2,\dots, 18\}$, there exists $w \in \{1,\dots, y-1\}$ such that $\M_{\STR\ph w}(1) > \M_{\SAR\ph y}(1)$ and $\ul{\T}_{\STR\ph w}(1 ; 0.27) < \ul{\T}_{\SAR\ph y}(1 ; 0.27)$).
}
This is not true in general.
Even when there exists $w$ such that $\STR\ph w$ better represents minorities than $\SAR\ph y$, $\STR\ph w$ might still exclude fewer extreme jurors than $\SAR\ph y$ if jurors are considered extreme when their conviction probability falls below an arbitrary $c > 0$.
However, an extension of Proposition \ref{prop:n_extreme} shows that when such a $w$ exists, there also exists $\ul{c} >0$ such that if jurors are considered extreme when their conviction probability falls below $\ul{c}$, $\STR\ph w$ performs better than $\SAR\ph y$ in terms of both objectives.

\begin{prop}
\label{prop:n_chall}
Consider any $x \in \{1,\dots, j\}$ and any $y \geq 1$. 
Suppose that there exists $w \geq 1$ such that $\M_{\STR\ph w}(x) > \M_{\SAR\ph y}(x)$.
Then for some $\ul{c} > 0$, we also have $\ul{\T}_{\STR\ph w}(x ; c) < \ul{\T}_{\SAR\ph y}(x ; c)$  for all $c \in (0, \ul{c})$.
\end{prop}

\section{Statistical Discrimination}
\label{sec:discrimination}

In our analysis, it was assumed that the defense and plaintiff were able to determine the jurors' probability of conviction with certainty. However, the accuracy of this probability estimation is limited by the incomplete information gathered during the \emph{voir dire} questioning process.
When the parties use group identity as an additional, albeit imperfect signal, statistical discrimination may occur. This is because jurors from different groups may be treated differently even when they exhibit the same signals of their attitudes toward conviction.\footnote{Statistical discrimination originates from  \cite{phelps1972, Arrow73}. For a survey, see \cite{fang2011theorie}.}
In this section, we discuss the extension of the results from Section \ref{sec:grouprep} when allowing for this incomplete information. 

Formally, let $c_{a}, c_{b}$ be the probability of a juror from group $a$ and $b$ casting a vote for conviction, respectively. The distributions of $c_{a}, c_{b}$, with densities $f_a, f_b$, are common knowledge, and so are the probabilities $r, 1-r$ of a juror drawing group identity $a, b$, respectively. In the model we have analyzed so far, which we label the \emph{baseline model}, we have assumed parties learn the conviction probability $c_i$ of each juror $i$ with certainty. In the \emph{Statistical discrimination model} instead, we assume parties observe the juror's group identity $g_i$, and a noisy signal $s_i$ of $c_i$, drawn from  a distribution with density $\theta(s|c_i)$. 
We assume that $\theta(s|c_1)/\theta(s|c_2)$ is strictly increasing in $s$ for all $c_1>c_2$. This monotone likelihood ratio property ensures that the signal is informative about the true probability of conviction: higher values of the signals are more likely when $c_i$ is higher.% %Necessary to avoid spacing with footnote
\footnote{In the parametric examples we propose later we characterize the signal distribution $\theta$ in terms of the signal's precision. For the purpose of our analysis, we assume that $\theta$ is given, neglecting the possibility that parties may strategically choose the signal's precision by prolonging their inquiry. In practice, judges do not allow \emph{voir dire} to continue indefinitely; hence our assumption is that signal precision is determined by the amount of time judges allocate for questioning. This assumption is reasonable as it is consistent with the practical consideration of the judge's time constraints.}

Parties use Bayes rule to compute the density of the probability of voting for conviction of juror $i$'s from group $g_i$, conditional on the signal and group identity:
\begin{equation}
f(c_i|s_i,g_i) = \frac{\theta(s_i|c_i)f_{g_i}(c_i)}{p_{g_i}(s_i)} \label{bayes}
\end{equation}
where $p_{g_i}(s_i)=\int_c\theta(s_i|c)f_{g_i}(c)dc$. Parties only care about group identity insofar as it is informative about $c_i$ from (\ref{bayes}). 

We denote this' density's expected value explicity as a function of the signal and group identity: $\xi(s_i;g_i) = E(c_i|s_i;g_i)$ Let $\Xi_g$ be the image of $\xi(\cdot,g)$, and $\Xi=\Xi_a \bigcup \Xi_b $, and denote with $q_a$ and  $q_b$ be the distributions over $\Xi_a,\Xi_b$ implied by $f_a, f_b,$ and $\theta$. 
The densities of $q_a, q_b$ are computed using the transformation of random variables rule applied to Bayes' formula denominator: 
$q_g(x)= p_{g}(\xi^{-1} (x,g))\left|\frac{d\xi^-1(x;g)}{dx}\right|$, $g=a,b$ (assuming $q_g$ is a continuous distribution).%
\footnote{In Appendix \ref{sec:posteriors} we characterize distributions $q_g$ formally for two parametric assumptions. In one of these parameterizations $q_g$ are discrete distributions, hence $q_g(x)= p_{g}(\xi^{-1} (x,g)), \forall x\in\Xi_g$.}
Note that the inverse $\xi^{-1}$ (over variable $s_i$) exists, because function $\xi$ is strictly increasing in $s_i$ from the monotone likelihood property assumption.

The proposition below states that the Statistical discrimination model is equivalent to the baseline model where parties observe jurors' true conviction probabilities taking values in $\Xi$ drawn from distributions with densities $q_a,q_b$. 

\begin{prop}\label{prop:statdisc}
    Let conviction probabilities $c$ of jurors from groups $a,b$ be drawn from distributions $C_a,C_b$ with densities $f_a, f_b$, let parties observe signals of $c$ drawn from density $\theta(s|c)$, satisfying a strict monotone likelihood ratio property, and let $q_a, q_b$ be the distributions (defined over support $\Xi_g \subseteq[0,1], g=a,b$) of the realized conditional expectations $E(c|s,g), g=a,b$ implied by $f_a, f_b, \theta$.
Then, in the model with statistical discrimination equilibrium strategies under selection rules \STR and \SAR coincide with the equilibrium strategies of the baseline model where parties learn conviction probabilities $c_i$ with certainty, and $c_i$ takes values in $\Xi_{g_i}$ drawn from distributions $q_{g_i}$, where $g_i$ is $i$'s group membership.
\end{prop}

Because the parties' problem is analogous to the problem presented in Section \ref{sec:model} results from the previous sections apply if the conditions supporting the results hold for distributions $q_a, q_b$. 
To gain some intuition about the validity of the previous sections' results under the statistical discrimination model, consider first the extreme case in which the signal is fully informative, that is $E(c_i|s_i,g_i)=c_i$ implying  $q(\xi(s_i,g_i)=c)=f_{g_i}(c), \forall c \in [0,1]$. The models are identical, all results from the previous sections apply, and by continuity, they will apply when signals are nearly fully informative, as we will confirm below using simulations. 

At the other extreme, the case where
there is no signal, parties only observe the prospective juror's group identity. Assume that groups are polarised with group $a$ favoring the defendant's acquittal, on average: $E(C_a)<E(C_b)$. The only signal of conviction probability in this case is group identity, and the expected conviction probability of every juror is their group average, therefore $q$'s support $\Xi$ will only take the values of the two group averages:
$$
q(c) = \begin{cases}
            r & \text{if } c=E(C_a)\\
            1-r & \text{if } c=E(C_b) \\
            0 & \text{otherwise}
        \end{cases}
$$
Regardless of the procedure, in such situation, $D$ will challenge all jurors from group $b$, and $P$ all jurors from group $a$. 
It is crucial that  $\Xi$ takes only two  values (the group averages), hence all jurors are ``extreme'': it is not possible for \SAR to run out of challenges before selecting a more extreme juror, or a juror of a different group is presented.\footnote{Propositions \ref{prop:n_extreme} and \ref{prop:minor_compare} do not apply because $q$ is not continuous, whereas the basic model assumes continuous distributions. Continuity, however, is not a necessary condition, as shown from the simulations in Appendix \ref{sec:posteriors}.} 
Later in this section, when discussing simulation results, we will revisit the concept of within-group heterogeneity. We will note that some degree of such heterogeneity is necessary to ensure that \SAR and \STR produce meaningful differences in outcomes.
%However, if $q(c)$ takes positive mass at at least three values of $c$, say $c_1<c_2<c_3$, then if $j>=2$ there are subgames with thresholds for $P$, $t_P>c_2$, then it is possible that \SAR runs out of challenges by challenging a juror with $E(c|s,g)=c_2$ and a more extreme juror with $c=c_1$ is selected. 

When the signal is only partially revealing, it is difficult to formally characterize whether statistical discrimination facilitates the exclusion of extremes and the inclusion of minorities under \SAR relative to \STR.
Simulations help provide some intuition. 

We parameterize signals according to their precision in revealing the true conviction probability and show that as the signal becomes more precise, the differences between \SAR and \STR become closer to the ones predicted by the baseline model, confirming that results from the baseline model are robust to adding statistical discrimination.

The parameterization we choose allows us to derive the conditional expectations of conviction probability in closed form.\footnote{We present this parameterization also because the posterior is a continuous random variable, and the propositions in the previous sections assume continuity. In Appendix \ref{sec:posteriors} we show results obtained from parameterizations with conviction probabilities drawn from Beta distributions, and signals arising from $N$ Bernoulli trials with parameter $p$ equal to jurors' true conviction probability. The posterior in this case is discrete, with $N+1$ points in its domain. The results from this parameterization are similar.}
We assume that the conviction probability follows a logit-normal distribution with parameters $(\mu, \sigma^2)$, i.e., a distribution whose logit transformation is a normal distribution with parameters $(\mu, \sigma^2)$.\footnote{If $Y$ is logit-normally distributed, then its logit $X=ln(y/(1-y))$ is normally distributed. Its domain is $[0,1]$ and if $\mu=0$ the distribution is symmetric around 0.5. See \cite{logit-normal}.}
In our simulations, we draw for each juror from group $g=a,b$ an auxiliary value $x$ from the Normal distribution $N(\mu_g,\sigma_g)$, and  assume the juror votes for conviction with probability $c=e^{x{}}/(1+e^{x{}})\in[0,1]\  \forall x\in \Re$ (we dropped subscript $i$ indicating the juror's identity for ease of notation).

In the statistical discrimination model we assume that parties do not observe a juror's $x$, nor its logistic transformation $c$, but observe a noisy signal $s=x+\varepsilon$, where $\varepsilon\sim N(0,\sigma_\varepsilon)$. We denote the signal precision with the inverse of its variance, $1/\sigma_\epsilon^2$. 
The posterior density of $x$ conditional on signal and group identity is itself normal, a result exploited in the seminal statistical discrimination literature (\cite{phelps1972}):
$$
f(x|s,g) \sim N\left(\alpha s + (1-\alpha)\mu_g, \alpha\sigma_\epsilon^2 \right) 
\text{ with } 
\alpha = \frac{\sigma^2}{\sigma^2+\sigma_\epsilon^2}
$$
which implies that the conditional expectation of $c$ is logit-normally distributed with parameters 
$(\alpha s + (1-\alpha)\mu_g, \alpha\sigma_\epsilon^2).$
Note that as precision increases ($\sigma_\epsilon\rightarrow 0$), $\alpha\rightarrow1 $ and $s\rightarrow x$, that is, the conditional expectation of $x$ converges to its true value. As precision decreases instead ($\sigma_\epsilon\rightarrow \infty$), $\alpha \rightarrow 0$, and $E(x|s,g)\rightarrow \mu_g$, the group average. The posterior density of the conditional expectations $q_g$ becomes degenerate, with only one value in its domain.

%\begin{figure}[t]
%    \centering
%    \caption{Minority representation in the Statistical Discrimination model}\label{fig:stdisc}
%    \begin{subfigure}{2.25in}
%    \begin{overpic}[width=2.25in]{std-logitnorm-2}
%     \put(12,40){\includegraphics[scale=0.25]{std-density-2.pdf}}  
%  \end{overpic}
%    \caption{Extreme polarization}\label{fig:stdbeta}
%    \end{subfigure}%
%    \begin{subfigure}{2.25in}
%    \begin{overpic}[width=2.25in]{std-logitnorm}
%     \put(12,40){\includegraphics[scale=0.25]{std-density-1.pdf}}  
%  \end{overpic}
%    \caption{Moderate  polarization}\label{fig:stdnormal}
%    \end{subfigure}
%    \caption*{\normalfont\footnotesize \emph{Notes:} Percent points differences between \SAR and \STR in the number of seated jurors from group $a$ (dashed lines) and in juries with at least one juror from group $a$ (solid lines), $r=0.25$, 50000 simulated juries per simulation. Panel (a):  $\mu_a=-2, \mu_b=2, \sigma_a=\sigma_b=1$. Panel (b): $\mu_a=-1, \mu_b=1, \sigma_a=\sigma_b=1$. Values corresponding to tick marker $\infty$ are computed using the baseline model. Each panel's inset displays the true distribution of conviction probabilities (brown solid line), each group's distributions (dashed lines), and the group averages (dotted vertical lines)}
%\end{figure}

\begin{figure}[t]
    \centering
    \caption{Minority representation in the Statistical Discrimination model}\label{fig:stdisc}
    \includegraphics[width=4.5in]{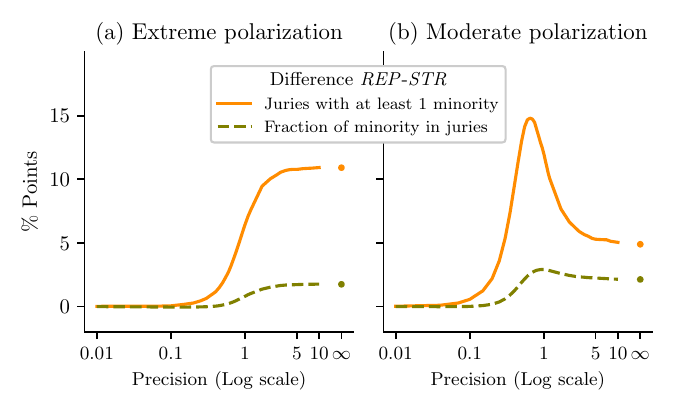}
\end{figure}

In Figure \ref{fig:stdisc} we report the results from two sets of simulations using this parameterization, using $j=12$, $d=p=6$, and fraction of minority $r=0.25$. 
The figure reports percent points differences between \SAR and \STR in the number of seated jurors from group $a$ (dashed lines) and in juries with at least one juror from group $a$ (solid lines).
The horizontal axis covers different values of signal precision. 

Simulations reported in Panel (a) adopt parameters implying a relatively large difference in the average conviction probability between groups ($\mu_a=-2, \mu_b=2, \sigma=1$). 
When the signal precision is very small parties essentially observe only group identity resulting in jurors being selected only based on their group; as a result, the two procedures do not differ substantially in their outcomes. At all precision levels, the \SAR procedure selects more minorities than \STR. As the signal precision increases, the difference between the two models converges towards the perfect information result. 
In the baseline model (denoted with precision $\infty$), \SAR selects more minorities than \STR with a 11 percentage point difference between the two models in the fraction of juries with at least one minority and a difference of approximately 2 percentage point in the fraction of seated minority jurors. 

With these parameters, differences in minority representation between models are smaller than in the baseline model at all levels of precision. This is not always the case.
When the true probability of conviction shows a more moderate degree of group polarization as in Panel (b) (drawn with parameters $\mu_a=-1, \mu_b=1,$ and all other parameters identical to the ones used in the previous simulation),  then 
intermediate values of precision produce a larger difference in minority representation between \SAR and \STR than under the baseline model. This difference reaches three times the baseline value of the fraction of juries with at least one minority when $\sigma_\epsilon^2$ is about 1 (solid line).   

The comparison between \SAR and \STR relies on how the signal's noise shifts the distribution of perceived conviction probabilities relative to the underlying true distributions. Two contrasting factors come into play: (i) the degree of overlap between the group distributions and (ii) the degree of within-group heterogeneity in conviction probability. 
As the signal noise increases, the posterior distributions start to separate and decrease in variance, eventually converging towards a degenerate distribution that places all mass at the group average. 
Proposition \ref{prop:minor_compare} states that \SAR selects more minorities than \STR when the two distributions do not overlap in the limit. Hence, it is possible for \SAR to select more minority jurors than \STR when only a relatively small portion of the distributions overlap (factor (i)).
However, the \emph{magnitude} of the difference also relies on the distribution's support to be sufficiently wide (factor (ii)):
there need to be subgames in \STR where jurors are challenged despite the possibility of more extreme jurors, or jurors from a different group, being drawn later in the game when the parties run out of challenges. As posterior distributions become sufficiently concentrated, the mass of jurors that are ``more extreme" than the threshold jurors shrinks. As we noted earlier, both procedures nearly select jurors on the basis of their race when there is minimal overlap between the distributions, and while Proposition \ref{prop:minor_compare} remains true, the difference between procedures vanishes and is zero at the limit.

\begin{figure}[t]
    \centering
    \caption{Conviction probability distributions used to generate Figure \ref{fig:stdisc}}\label{fig:stdiscdistros}
    \includegraphics[width=4.5in]{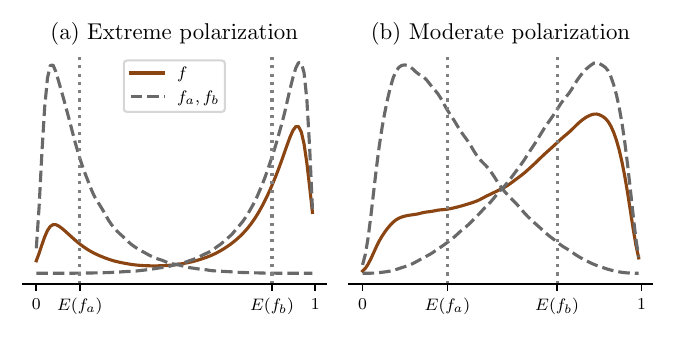}
\end{figure}

The true distributions of conviction probabilities used to generate then simulations from Figure \ref{fig:stdisc} are displayed in Figure \ref{fig:stdiscdistros}. Notice the true group distributions (dashed lines) of $c$ have a large overlap area in the simulation with moderate polarization (panel (b)). As noise increases, the overlap area decreases, therefore initially the proposition's assumption is more easily satisfied, and outcomes differences between models increase. With the parameters adopted for the simulations of Panel (a), the true group distributions overlap very little and are relatively more concentrated around the group average (vertical dotted lines), hence factor (ii) prevails even at high levels of precision.  

In summary, this section demonstrates that the conclusions from the previous sections remain valid even when conviction probabilities are observed with noise unless the signal is completely uninformative. Specifically, when groups are sufficiently polarized, the \SAR model selects a higher proportion of minorities than the \STR model, but under statistical discrimination, differences between models, as measured by specific outcomes, are generally smaller than in the baseline model, with exceptions occurring for some parameters at an intermediate level of the signal precision.

\section{Extensions}%: Unbalanced juries and representation of balanced groups}
\label{sec:ext}
%%%%%%%%%%%%%%%%%%%%%%%%
%%%%%%%%%%%%%%%%%%%%%%%%

In this Section we consider additional extensions of the baseline model. 

%%%%%%%%%%%%%%%%%%%%%%%%
\subsection{Excluding unbalanced juries} \label{sec:unbalanced}
%%%%%%%%%%%%%%%%%%%%%%%%

The primary purpose of jury selection is to prevent extreme jurors from serving (see Footnote \ref{foot:partial}).
In our model, it seems natural to interpret this goal as that of limiting the selection of jurors coming from the tail of the distribution, as we have done so far.
Another approach is to consider the extremism of juries \emph{as a whole}.
%There are different metrics one can use to evaluate how effective a procedure is at excluding extreme juries. 
%%If extreme juries are viewed as juries composed jurors with positions above and below given thresholds, then Propositions \ref{prop:n_extreme} and \ref{prop:extremeRAN} suggest that $\STR$ is more effective than both \SAR and \RAN\ excluding extreme juries.
For example, extreme juries could  be juries in which the juror with the highest or lowest conviction probability is extreme.
Using variants of the arguments in the proofs of Propositions \ref{prop:n_extreme} and \ref{prop:extremeRAN}, one can show that, in that sense too, $\STR$ is more effective than both \SAR and \RAN\ at excluding extreme juries.\footnote{
Specifically, for any $x \in \{0,\dots, j-1\}$,
	there exists $\ul{c} > 0$ and $\bar{c} < 1$, such that 
	(a) for every  $c \in (0, \ul{c})$, the probability that the lowest conviction-probability in the jury is smaller than $c$ is larger under \SAR and \RAN\ than under $\STR$, and
	(b) for every $c \in (\bar{c},1)$, the probability that the highest conviction-probability in the jury is larger than $c$ is larger under \SAR and \RAN\ than under $\STR$.
}

Another measure of juries' extremism, proposed by \cite{flanagan_peremptory_2015}, is whether a jury is excessively ``unbalanced" in the sense of featuring a disproportionate proportion of jurors coming from one side of the median of $C$.
Interestingly, Flanagan shows that \STR introduces correlation between the selected jurors, which leads the procedure to select more unbalanced juries than \RAN.
Even though panels are the result of independent draws from the population, jurors selected under \STR have conviction probabilities between that of the lowest and highest challenged juror.
For example, the selection of two jurors with conviction probabilities $0.25$ and $0.75$ indicates that challenges were used on jurors with conviction probabilities outside the $[0.25, 0.75]$ range.
The latter makes it more likely that \STR selected additional jurors between $[0.25, 0.75]$, introducing a correlation between selected jurors.\footnote{This intuition is one of the main points in \cite{flanagan_peremptory_2015}. In Corollary 2 he  shows that, even when the parties have the same number of challenges ($d = p$), the probability that \emph{all} selected jurors come from one side of the median is \emph{larger} under $\STR$ than under \RAN. 
Our next proposition, using a new proof technique, generalizes this result, for any number of jurors larger than one have the jury size (not just $x=j$).}

Formalizing this intuition, we show that for \emph{any} $x$ larger than half the jury-size, the probability of selecting at least $x$ jurors from one side of the median is larger under \STR than under \RAN.
As in Section \ref{sec:extreme}, we focus on the probability that the selected jurors are \emph{below} the median (our results apply symmetrically to selection above the median).
Let $med[C]$ denote the median of $C$.

\begin{prop}
\label{prop:median}
%  (a) 
 If $d = p$, then for any $x \in \{j/2 + 1, \dots, j\}$ if $j$ is even, and any $x \in \{j/2 + 1.5, \dots, j\}$ if $j$ is odd, we have $\ul{\T}_{\STR}\big(x ; med[C]\big) > \ul{\T}_{\RAN}\big(x ; med[C]\big)$.
%  
%  In contrast, 
%  (b) for different values of $j$, $d = p$, and $x \in \{\ceil{(j+1)/2}, \dots, j\}$ and different distributions $f(c)$ it is possible to have either $\ul{\T}_{\SAR}(x ; med[C]) > \ul{\T}_{\STR}(x ; med[C])$, 
%  or
%  $\ul{\T}_{\SAR}(x ; med[C]) < \ul{\T}_{\RAN}(x ; med[C]) < \ul{\T}_{\STR}(x ; med[C])$.
	\end{prop}

\begin{figure}[t!]
\centering
\caption{Selection of jurors below the median
\label{fig:median}}
\includegraphics[width = 4in]{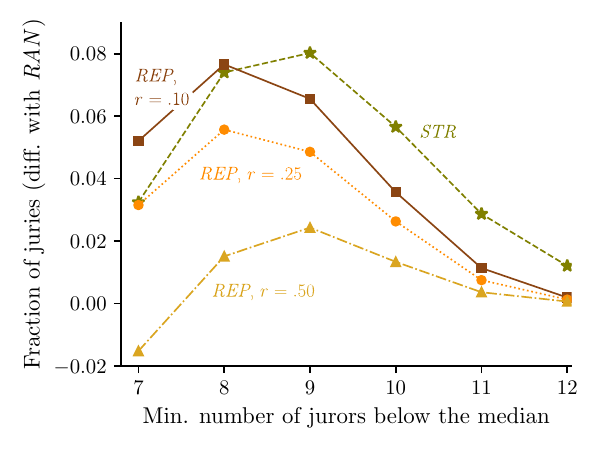}

%\caption*{\footnotesize \normalfont \emph{Note:}  Fraction of juries with a at least given number of jurors below the median of $C$  under \STR (green dashed line) and \SAR (continuous lines) relative to the same fraction under \RAN (i.e.  $\ul{\T}_{M}(x ; med[C]) - \ul{\T}_{\RAN}(x ; med[C])$). 
%Throughout, we fix $j=12$, $d=p = 6$ and $C\sim r * Beta(1,5)+(1-r) * Beta(5,1)$ (for $r \in \{0.1,0.25,0.5\}$) whereas the number of jurors below the median is on the horizontal axis.  
%For each set of parameters, results for \SAR are averages across 50,000 simulated jury selections, whereas values for \RAN and \STR are computed analytically and are independent of $r$ (see Footnote \ref{foot:analytical}).
%}
\end{figure}

For $M \in \{\STR,\RAN\}$, the value of $\ul{\T}_{M}\left(x ; med[C]\right)$ can be computed analytically and does not depend on the distribution of $C$.\footnote{\label{foot:analytical}%
Specifically,
$\ul{\T}_{\RAN}\big(x ; med[C]\big) = \p(Bi[j,0.5] \geq x)$ and
$\ul{\T}_{\STR}\big(x ; med[C]\big) = \p(Bi[j + d + p, 0.5] \geq x + p)$.
}
For $M=\SAR$, the value of $\ul{\T}_{M}\left(x ; med[C]\right)$ depends on the distribution in a complex fashion and it is not possible to generally compare $\SAR$ with the two other procedures in terms of $\ul{\T}_{M}\left(x ; med[C]\right)$.
In Figure \ref{fig:median} we illustrate Proposition \ref{prop:median}, and that a similar statement does not hold for \SAR, simulating outcomes using $j=12$, $d=p = 6$ and $C\sim r * Beta(1,5)+(1-r) * Beta(5,1)$ (for $r \in \{0.1,0.25,0.5\}$).
As the figure shows, the fraction of simulated juries with at least $x$ jurors below $med[C]$ can, in some cases (in the figure, $x=7$ and, barely, $x=8$ jurors), be larger under \SAR than under both \RAN and \STR.
In other cases, however, the same figure is lower under \SAR than under both \RAN and \STR.

% For example, for the case where $j=12, d=p=6$, we have 
% $\ol{\T}_{RAN}\left(7 ; med[C]\right) = .39 >
%  \ol{\T}_{STR}\left(7 ; med[C]\right) = .42 
% $ independently on the distribution and consistently with Proposition \ref{prop:median}. 
% Under the distribution used to construct Figure \ref{fig:atleast1-3betas}, we obtain from simulations $\ol{\T}_{\SAR}\left(7 ; med[C]\right) = .43$ that is, greater than the values of the same statistic under \STR and \RAN. 
% However, with more left-skewed distributions we obtain values where \SAR is less likely to select at least 7 jurors are above the median than either \STR, or both \STR and \RAN. 
% Using 
% $f(c)= 0.25* Beta(c;\alpha=2,\beta=4) + 0.75* Beta(c;\alpha=4,\beta=2)$ we obtain $\ol{\T}_{\SAR}\left(7 ; med[C]\right) = .40$, 
% and with 
% $f(c)= 0.1* Beta(c;\alpha=2,\beta=4) + 0.9* Beta(c;\alpha=4,\beta=2)$ we obtain $\ol{\T}_{\SAR}\left(7 ; med[C]\right) = .38$.

% Increasing skewness even further increases $\ol{\T}_{\SAR}\left(7 ; med[C]\right)$ to values above $\ol{\T}_{\STR}\left(7 ; med[C]\right)$, therefore we could not find any clear pattern in the comparison between \STR and \SAR. 

Figure \ref{fig:median} displays the result of simulations when the distribution of $C$ is highly polarized (a mixture of $Beta(1,5)$ and $Beta(5,1)$).
In the online appendix (\cite{mvdl-external}) we present additional simulations for less polarized distributions.
These additional simulations suggest that high levels of polarization are required for \SAR to more often select a majority of jurors below the median than \STR.
Also, for lower levels of polarization, \SAR tends to selects fewer juries made of a majority of jurors below the median than \RAN.\footnote{
Because the parties' actions under \SAR are influenced by the mean of the distribution but not in any clear way by the median (and because of the complexity of the game tree), we were unable to formalize the effect of polarization on these comparisons in terms of the model parameters. 
}

%%%%%%%%%%%%%%%%%%%%%%%%
\subsection{Representation of balanced groups}
%%%%%%%%%%%%%%%%%%%%%%%%

\label{sec:balanced}

Even though the U.S. Supreme Court initially banned challenges based on race only  (\emph{Batson v. Kentucky}, 1986), it later  banned challenges based on \emph{gender} (\emph{J.E.B. v. Alabama}, 1994).
It is therefore natural to ask whether the advantage of \SAR in terms of minority representation comes at the cost of a worse representation of gender groups.

Unlike minorities which correspond to groups of unequal sizes represented by small values of $r$, gender-groups can be thought of as even-sized groups and are better modeled using $r \approx 0.5$.
With groups of similar sizes, both procedures almost always select at least a few members from either group.
It is therefore more interesting to compare procedures directly in terms of the \emph{proportion} of group-$a$ jurors they select (rather than in terms of the probability of selecting \emph{at least} $x$ members from group-$a$, as we did before).

In this last section, we let $r = 0.5$ and study the expected proportion of group-$a$ jurors selected under \STR and \SAR.
We denote these proportions $r_{\STR}$ and $r_{\SAR}$ and focus on how close $r_{\STR}$ and $r_{\SAR}$ are from the 50\% of group-$a$ jurors that prevail in the population.
% \footnote{
% Previous results are stronger in the sense that they establish a first-order stochastic dominance between the number of jurors with certain characteristics  (extremism or group-membership) selected under \STR and \SAR.
% As we explain after Proposition \ref{prop:n_extreme}, showing, for example, that $\ul{\T}_{STR}(x ; c) < \ul{\T}_{\SAR}(x ; c)$ for all $x \in \{1, \dots, j\}$ directly implies that the expected proportion of selected jurors with conviction probability $c_i < c$ is lower under $\STR$ than under $\SAR$ (whereas the converse is not true).
% }

As in the last two sections, it is not possible to generally compare \STR and \SAR in terms of the procedures' ability to select an even proportion of group-$a$ and group-$b$ jurors. 
In some cases, $r_{\STR}$ can be further away from 50\% than $r_{\SAR}$, and the converse may be true in other cases.
For example, with $d = p = 6$ and $j = 12$, if $C_a \sim U[0,1]$ and $C_b \sim Beta(1,5)$, simulations reveal that $r_{\STR} = 43.7\%$ whereas $r_{\SAR} = 45.8 \%$.
In contrast, when $C_a \sim Beta[4,2]$ and $C_b \sim Beta(1,5)$, $r_{\STR} = 50.3\%$ whereas $r_{\SAR} = 52.2 \%$.

These examples however suggest that, as joint distribution becomes more symmetrical, $r_\STR$ get closer to 50\% .
Proposition \ref{prop:compare_sym} confirms this pattern.
If the group-distributions exibit mirror-symmetry (or if they do not overlap) and if $d = p$, then  $r_{\STR} = 50\%$ whereas \SAR does not necessarily select an even proportion of jurors from each group.
This is because even when $r= 50\%$ and distributions are symmetrical, the multiplicative utility function that the parties use to assess the value of a jury (a consequence of the assumption that convictions require unanimity) creates asymmetries in the use of challenges under \SAR.%
\footnote{\cite{flanagan_peremptory_2015} shows that, in this symmetrical case, the asymmetry of the payoffs still forces the defendant to be more conservative than the plaintiff when using its challenges, hence leading to an uneven selection of jurors from the two groups.}

We say that random variables $C_a$ and $C_b$ exhibit \textbf{mirror-symmetry} if $f_a(c) = f_b(1-c), \forall c \in [0,1]$. 

\begin{prop} \label{prop:compare_sym}
    Suppose that $r = 0.5$ and $d = p$.
    If
    (a) the two group distributions do not overlap,\footnote{
    That is either $\p(C_a > C_b) = 0$ or $\p(C_b > C_a) = 0$.
    The same result would apply if the two distributions did not overlap \emph{in the limit as in Proposition \ref{prop:minor_compare}.}
    } or 
    (b) $C_a$ and $C_b$ exhibit mirror-symmetry, then
    $r_{STR} = r_{RAN}$.
\end{prop}

\begin{figure}[t!]
    \centering
    \caption{Representation of Group-a jurors with balanced group sizes}\label{tab:betas-grouprep-balanced}
\includegraphics[width=4.5in]{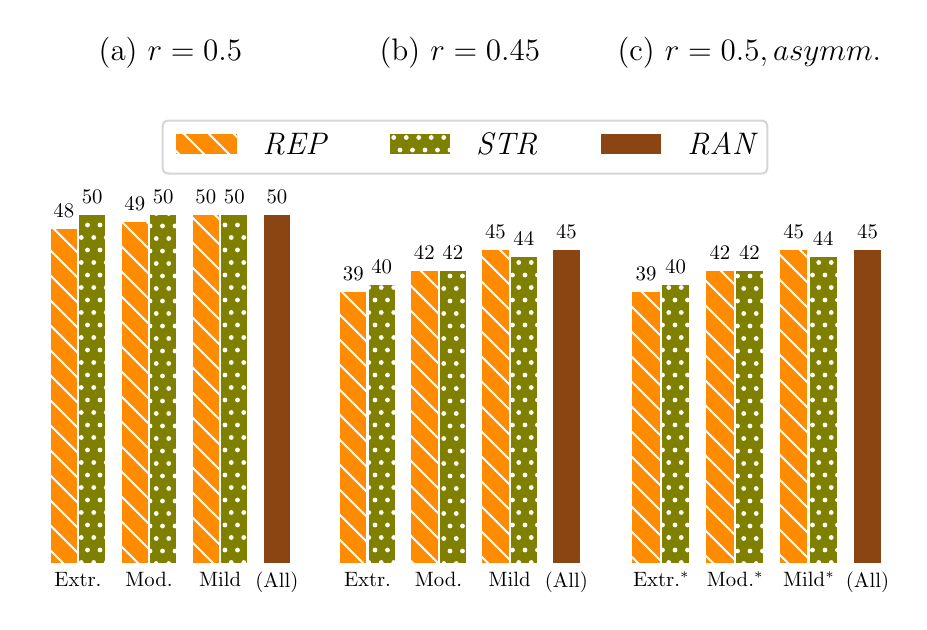}
%
%    \caption*{\normalfont\footnotesize \emph{Notes:} Percent group-$a$ seated jurors out of 50,000 simulations of jury selection with parameters $j=12$ and $d=p=6$. $^{*}$In panel (c) Extreme$^{*}$ corresponds to $C_a \sim Beta(1,5)$ and $C_b \sim Beta(5,2)$, 
%    Moderate$^{*}$ to $C_a \sim Beta(2,4)$ and $C_b \sim Beta(4,3)$, and
%	Mild$^{*}$ to $C_a \sim Beta(3,4)$ and $C_b \sim Beta(4,4)$.
\end{figure}

Figure \ref{tab:betas-grouprep-balanced}(a) illustrates Proposition \ref{prop:compare_sym} and the fact that a sixmilar statement does not hold for \SAR. The simulations are computed using the distributions displayed in Figure Figure \ref{fig:betaPDFs}, and $r=0.5$. 
Unlike \STR, \SAR can select unequal numbers of group-$a$ and group-$b$ jurors even when the joint distribution is symmetric around 0.5 (but groups are polarized).
Therefore, as a consequence of Proposition \ref{prop:compare_sym}, $r_\SAR$ can in these cases be further away than $r_\STR$ from the 50\% of group-$a$ jurors that prevail in the population.

Figure \ref{tab:betas-grouprep-balanced}(a) however suggests that these differences may be quantitatively small, and that sizable differences may require high levels of polarization between groups.
Figure \ref{tab:betas-grouprep-balanced}(b) and \ref{tab:betas-grouprep-balanced}(c) report the results of simulations in which the symmetries required for Proposition \ref{prop:compare_sym} to hold are slightly relaxed.
These indicate that the advantage of \STR in the representation of balanced groups established in Proposition \ref{prop:compare_sym} (i.e., the fact that $r_\STR$ is closer to 50\% than $r_\SAR$) may not be robust to even mild relaxations of these symmetries.
In particular, when $r = 0.45$ (Figure \ref{tab:betas-grouprep-balanced}(b)) or when $r = 0.5$ but the group-distributions exhibit a slight mirror-asymmetry (Figure \ref{tab:betas-grouprep-balanced}(c))\footnote{In panel (c) Extreme$^{*}$ corresponds to $C_a \sim Beta(1,5)$ and $C_b \sim Beta(5,2)$, 
Moderate$^{*}$ to $C_a \sim Beta(2,4)$ and $C_b \sim Beta(4,3)$, and
Mild$^{*}$ to $C_a \sim Beta(3,4)$ and $C_b \sim Beta(4,4)$}, 
$r_\SAR$ is closer than $r_\STR$ to the proportion of group-$a$ jurors that prevail in the population for some levels of polarization.

\section{Conclusion} 
In this paper, we study the relative performance of two stylized jury-selection procedures. Strike and Replace presents potential jurors one-by-one  to  the  parties, whereas the  Struck  procedure   presents  all  potential  jurors  before they exercise vetoes. When jurors differ in their probability of voting for the defendant's conviction, and belong to polarized groups, Struck is more effective at excluding jurors with extreme views, but generally selects fewer members of a minority than Strike and Replace, leading to a conflict between these goals.

The legal debate emphasizes the effect of the choice of jury selection procedure in selecting different types of jurors, motivating the main goal of our analysis. 
Another important topic is the procedures' differential effect on trial \emph{outcomes}, for example, the prediction of which procedure is more likely to lead to conviction or to a \emph{correct} verdict. Studying these questions requires additional assumptions about actual jury behavior (not only about the parties' expectations of jury deliberations) and further departures from existing models of jury selection. We leave the analysis of these important questions to future research.

 Besides the selection of juries, this research may be suggestive of applications to other contexts where the mechanisms or procedures used to select (groups of) agents may have disparate outcomes on group-representation.
 One example is the voting rules that hiring committees use to select job candidates for interviews and fly-outs.
 
Sociologists \cite{small-pager-2020} argue that systemic factors may lead to disparate outcomes even in the absence of taste-based or statistical discrimination, the traditional explanations for group inequalities in Economics. 
In our model, it is natural for asymmetric group preferences to generate asymmetric outcomes. Our results emphasize that the chosen selection procedure may exacerbate such asymmetries. 
This paper formalizes an example in which the pursuit of one objective, preventing extreme jurors to serve on juries, may lead to larger group disparities even if mechanisms and institutions are formally race-neutral. 

% \martinLi{We can leave it like this for now and see how people react. 
% Just wanted to point out that our model could be isomorphically reinterpreted as featuring statistical discrimination.
% Right now, we assume that whenever the parties observe a juror $i$, they observe its probability of voting of conviction $c_i$.
% More likely, what they do is observe the jurors' characteristics $X_i$, including group-membership $G_i$, and infer an expected probability of voting of conviction $E(X_i, G_i)$ on which they base their challenge choices (the model then carries on identically based on $E(X_i, G_i)$'s instead of $c_i$'s).
% To the extent that $G_i$ is not fully subsumed in the other $X_i$, you could argue that our parties do rely on some form of statistical discrimination.} 

% An empirical evaluation of this trade-off is outside the scope of this paper, but could shed light on the relative effectiveness of existing procedure or peremptory challenge reforms in pursuing these two important goals. 

%%%%%%%%%%%%%%%%%%%%%%%%%%%%%%%%%%%%
%%%%%%%%%%%%%%%%%%%%%%%%%%%%%%%%%%%%
%%%%%%%%%%%%%%%%%%%%%%%%%%%%%%%%%%%%
%%%%%%%%%%%%%%%%%%%%%%%%%%%%%%%%%%%%

\appendix

%\renewcommand\thefigure{\thesection.\arabic{figure}}
%\renewcommand{\thetable}{\thetable.\arabic{table}}
%%%%%%%%%%%%%%%%%%%%%%%%
%%%%%%%%%%%%%%%%%%%%%%%%
%%%%%%%%%%%%%%%%%%%%%%%%
\section{Appendix}

\setcounter{equation}{0}
 \renewcommand{\theequation}{\thesection.\arabic{equation}}

%%%%%%%%%%%%%%%%%%%%%%%%
%%%%%%%%%%%%%%%%%%%%%%%%
%%%%%%%%%%%%%%%%%%%%%%%%

%%%%%%%%%%%%%%%%%%%%%%%%
%%%%%%%%%%%%%%%%%%%%%%%%
\subsection{Preliminary technical results needed for proofs}\label{sec:prelimresults}
%%%%%%%%%%%%%%%%%%%%%%%%
%%%%%%%%%%%%%%%%%%%%%%%%

%%%%%%%%%%%%%%%%%%%%%%%%
\subsubsection*{Limit of a ratio of binomial probabilities}
%%%%%%%%%%%%%%%%%%%%%%%%

	\begin{lem}
 	\label{lem:binom_gen}
 	
 	For all $\eta \in \mathbb{N}$ and any $k \in \{1, \dots, \eta-1\}$, 
 	$$\lim_{\pi\rightarrow 0}~ \frac{\p[Bi(\eta,\pi) = k]}{\p[Bi(\eta,\pi) > k]} = \infty.$$
 	
 	\begin{proof}
 		\renewcommand{\qedsymbol}{$\square$}
 		Using the standard formula for the p.d.f. of a binomial and the representation of the c.d.f. of the binomial with \emph{regularized incomplete beta function}, we can re-write the ratio as
 		\begin{equation}
 		\label{eq:incomp_beta}
 		\frac{\p[Bi(\eta,\pi) = k]}{1 - \p[Bi(\eta,\pi) \leq k]} = 
 		\frac{{\binom{\eta}{k}
 	%	\eta\choose k
 		} \pi^k (1-\pi)^{\eta-k}}{1 - (\eta-k) 
 		\binom{\eta}{k}
 		%\binom{\eta}{k}
 		\int_0^{1-\pi} x^{\eta-k-1} (1-x)^k dx }
 		\end{equation}
 			As $\pi\rightarrow 0$, both the numerator and the denominator tend to $0$. 
 			We use L'Hopital's rule to complete the proof:
 		$$ \frac{(\partial/\partial \pi) ~ \binom{\eta}{k} \pi^k (1-\pi)^{\eta-k}}{
 			(\partial/\partial \pi) ~ \left(1 - \left[(\eta-k) 
 			\binom{\eta}{k}
 			%{\eta\choose r}
 			\int_0^{1-\pi} x^{\eta-k-1} (1-x)^k dx \right] \right) }$$
 		
 		$$= \frac{\binom{\eta}{k} * \left[ k\pi^{k-1} (1-\pi)^{\eta-k} + \pi^k (\eta-k) (1-\pi)^{\eta-k-1}\right]}{- (\eta-k)\binom{\eta}{k} \left[ (-1)  * (1-\pi)^{\eta-k-1} \pi^k  \right]} $$
 		
 		$$= \frac{ k\pi^{k-1} (1-\pi)^{\eta-k}}
 		{(\eta-k)    (1-\pi)^{\eta-k-1} \pi^k}  + \frac{ \pi^k (\eta-k) (1-\pi)^{\eta-k-1}}
 		{(\eta-k)    (1-\pi)^{\eta-k-1} \pi^k}$$
 		
 		$$= \frac{ k (1-\pi)}
 		{(\eta-k)     \pi}  + 1  ~~ \xrightarrow[\pi \rightarrow 0]{} ~~ \infty$$
 	\end{proof}		
 \end{lem}
 
%%%%%%%%%%%%%%%%%%%%%%%%
\subsubsection*{Continuity of challenge thresholds in \SAR as \texorpdfstring{$C^i$}{Ci} converges in distribution}
%%%%%%%%%%%%%%%%%%%%%%%%

\begin{lem}
	\label{lem:converg}
	Consider a sequence of random variables $\{C^i\}_{i = 1}^\infty$ that converges in distribution to some random variable $C^*$.
	Let ${t}_I(\gamma, C^i\big)$ denote the challenge threshold used by party $I \in \{D,P\}$ in an arbitrary subgame $\gamma$ of \SAR when the distribution of conviction probabilities is $C^i$.
	For any such subgame $\gamma$, we have $\lim_{i \ra \infty} {t}_I(\gamma, C^i\big) = {t}_I(\gamma, C^*)$. 
	
	%	Let $\T^i_{M}(x;c)$ and $\T^*_{M}(x;c)$ denote the probabilities that procedure $M$ selects at least $x$ jurors with conviction type smaller or equal to $c$ when the distribution of conviction probability in the population are $C^i$ and $C^*$, respectively.
	%	For both $M \in \{\STR, \SAR\}$, we have $\lim_{i\ra \infty} \T^i_{M}(x;c) = \T^*_{M}(x;c)$.
	
	\begin{proof}
		In any subgame $\tilde{\gamma}$, ${t}_I(\tilde{\gamma}, C^i\big)$ is the ratio of the value of continuation subgames if $I$ challenges the presented juror, or if both parties abstain from challenging \citep{brams_optimal_1978}.
		Therefore, $\lim_{i \ra \infty} {t}_I(\gamma, C^i\big) = {t}_I(\gamma, C^*)$ follows directly if we show that the value of any subgame, which we denote $V(\gamma, C^i\big)$, converges to $V(\gamma, C^*)$ as $i$ tends to infinity.\footnote{
			Because we assume that all distributions of conviction probabilities are continuous, there are no issues related to the possibility for the bottom of one of these ratios to converge to zero.
		} 
		
		The latter follows directly from the recursive characterization of $V(\gamma, C^i\big)$ in \cite{brams_optimal_1978}.
		Recall that each subgame $\gamma$ can be characterized by 
		the number of jurors $\kappa$ that remain to be selected, 
		the number of challenges left to the defendant $\delta$, and 
		the number of challenges left to the plaintiff $\pi$.
		With this notation, the recursive proof that for all $\kappa, \delta, \pi \geq 0$,  $V\big([\kappa, \delta, \pi], C^i\big)$ converges to $V\big([\kappa, \delta, \pi], C^*\big)$ as $i$ tends to infinity can be decomposed in a number of cases.
		Let $F^i(c)$ denote the the c.d.f. of $C^i$, $F^*(c)$ the c.d.f. of $C^*$, and $F(c)$ the c.d.f. of an arbitrary distribution $C$, with $\mu^i$, $\mu^*$, and $\mu$ being the corresponding expected values.
		In each step, the initial formula for $V\big([\kappa, \delta, \pi], C^i\big)$ is taken from \cite{brams_optimal_1978}.
		
		%%%%%%%%%%%%%%%%
		\medskip
		\textbf{Case 1: $\kappa = 0, \delta \geq 0, \pi \geq 0$.}
		%%%%%%%%%%%%%%%%
		In this case, $V\big([0, \delta, \pi], C) = 1$ for all $C$ and the convergence of $V\big([0, \delta, \pi], C^i\big)$ to $V\big([0, \delta, \pi], C^*)$ follows trivially.
		
		%%%%%%%%%%%%%%%%
		\medskip
		\textbf{Case 2: $\kappa > 0, \delta = 0, \pi = 0$.}
		%%%%%%%%%%%%%%%%
		In this case, $V\big([\kappa, 0, 0], C) = \mu^\kappa$ for all $C$ and the convergence of $V\big([0, \delta, \pi], C^i\big)$ to $V\big([0, \delta, \pi], C^*)$ follows from the fact that $C^i$ converges in distribution to $C^*$.
		
		%%%%%%%%%%%%%%%%
		\medskip
		\textbf{Case 3: $\kappa > 0, \delta = 0, \pi > 0$.}
		%%%%%%%%%%%%%%%%
		In this case, for all $C$,
		$$V\big([\kappa, 0, \pi], C) = V(\kappa - 1, 0, \pi)* \left[1 - \int_{t_I([\kappa, 0, \pi], C)}^1 F(c)~ dc \right],$$ 
		and $t_I([\kappa, 0, \pi], C) = V\big([\kappa, 0, \pi-1], C)/V\big([\kappa-1, 0, \pi], C)$.
		The convergence of $V\big([\kappa, 0, \pi], C^i\big)$ to $V\big([\kappa, 0, \pi], C^*)$ then follows recursively from the previous cases and from $C^i$ converging in distribution to $C^*$.

		%%%%%%%%%%%%%%%%
		\medskip
		\textbf{Case 4: $\kappa > 0, \delta > 0, \pi = 0$.}
		%%%%%%%%%%%%%%%%
		In this case, for all $C$,
		$$V\big([\kappa, \delta, 0], C) = V\big([\kappa, \delta - 1, 0], C) - V\big([\kappa-1, \delta, 0], C) * \int_0^{t_D([\kappa, \delta, 0], C)} F(c)~ dc,$$ 
		where $t_D([\kappa, \delta, 0], C) = V\big([\kappa, \delta - 1, 0], C)/V\big([\kappa-1, \delta, 0], C)$. 
		The convergence of $V\big([\kappa, \delta, \pi], C^i\big)$ to $V\big([\kappa, \delta, \pi], C^*)$ then follows recursively from the previous cases and from  $C^i$ converging in distribution to $C^*$.

		%%%%%%%%%%%%%%%%
		\medskip
		\textbf{Case 5: $\kappa > 0, \delta > 0, \pi > 0$.}
		%%%%%%%%%%%%%%%%
		In this case, for all $C$,
		$$V\big([\kappa, \delta, \pi], C) = V\big([\kappa, \delta - 1, \pi], C) - V\big([\kappa-1, \delta, \pi], C) * \int_{t_I([\kappa, \delta, \pi], C)}^{t_D([\kappa, \delta, \pi], C)} F(c)~ dc,$$ 
		where $t_D([\kappa, \delta, \pi], C) = V\big([\kappa, \delta - 1, \pi], C)/V\big([\kappa-1, \delta, \pi], C)$ and and $t_I([\kappa, \delta, \pi], C) = V\big([\kappa, \delta, \pi-1], C)/V\big([\kappa-1, \delta, \pi], C)$.
		The convergence of $V\big([\kappa, \delta, 0], C^i\big)$ to $V\big([\kappa, \delta, 0], C^*)$ follows recursively from the previous cases and from  $C^i$ converging in distribution to $C^*$.
%
		%		%%%%%
		%		\textbf{M = \STR.}
		%		%%%%%
		%		
		%		
		%		
		%		%%%%%
		%		\medskip
		%		\textbf{M = \SAR.}
		%		%%%%%
		%		
	\end{proof}
\end{lem}

%%%%%%%%%%%%%%%%%%%%%%%%
\subsubsection*{Comparative statics of probabilities from a symmetric binomial}
%%%%%%%%%%%%%%%%%%%%%%%%

\begin{lem}
\label{lem:comp_stat}
 ${\p[Bi(\eta +2,0.5) \geq k+1]} > {\p[Bi(\eta,0.5) \geq k]}$ \emph{if and only if} $k > \frac{\eta}{2} + \frac{1}{2}$.
 
 \begin{proof}
 We can decompose $\p[Bi(\eta +2,0.5) \geq k+1]$ in terms of $Bi(\eta,0.5)$ and $Bi(2,0.5)$:
 \begin{align*}
      & \p[Bi(\eta +2,0.5) \geq k+1]   \\
    = ~~~ & \p[Bi(\eta,0.5) \geq k+1] ~+~ 
     \p[Bi(\eta,0.5) =  k]  ~*~  \p[Bi(2,0.5)  \geq  1] ~+ \\
     & \p[Bi(\eta,0.5) =  k-1]  ~*~  \p[Bi(2,0.5)  =  2]  \\
     = ~~~ & \p[Bi(\eta,0.5) \geq k+1] ~+~ 
     \p[Bi(\eta,0.5) =  k]~*~ 0.75 ~+~
     \p[Bi(\eta,0.5) =  k-1]~*~ 0.25
 \end{align*}
 Also, 
 $${\p[Bi(\eta,0.5) \geq k]} = {\p[Bi(\eta,0.5) \geq k+1]} + \p[Bi(\eta,0.5) = k].$$
 The last two equalities imply that $\p[Bi(\eta +2,0.5) \geq k+1] ~>~ \p[Bi(\eta,0.5) \geq k]$ iff
 \begin{eqnarray*}
     \p[Bi(\eta,0.5) =  k]* 0.75 +
     \p[Bi(\eta,0.5) =  k-1]* 0.25 &>&  \p[Bi(\eta,0.5) = k] \\
     \p[Bi(\eta,0.5) =  k-1]* 0.25 &>&  \p[Bi(\eta,0.5) = k] * 0.25 \\
     \p[Bi(\eta,0.5) =  k-1] &>& \p[Bi(\eta,0.5) = k] \\
     %{\eta \choose k-1}
     \binom{\eta}{k-1}
     0.5^{k-1}0.5^{\eta- (k-1)} &>& 
     \binom{\eta}{k}
     %{\eta \choose k} 
     0.5^{k}0.5^{\eta- k} \\ 
     \frac{\eta!}{(\eta-[k-1])!(k-1)!} &>& \frac{\eta!}{(\eta-k)!k!} \\
     \frac{(\eta-k)!}{(\eta-[k-1])!} &>& \frac{(k-1)!}{k!} \\
     \frac{1}{\eta-k+1} &>& \frac{1}{k} \\ 
      k &>& \frac{\eta}{2} + \frac{1}{2}
 \end{eqnarray*}

 \end{proof}
 
\end{lem}

%%%%%%%%%%%%%%%%%%%%%%%%%%%%%%%%%%%% 
\subsubsection*{Relationship between order statistics of symmetric distributions}
%%%%%%%%%%%%%%%%%%%%%%%%%%%%%%%%%%%% 

For any number of draws $w$ and any $k\leq w$, let $C^{k,w}_g$ denote the $k$-th order statistic out of $w$ draws from distribution $C_g$, and $f^{k,w}_g(x)$ the corresponding probability density function.

\begin{lem}
	\label{lem:order}
	Suppose that $C_a$ and $C_b$ are symmetric.
	Then, for any $w \in \mathbb{N}$ and any $k \in \{1, \dots, w\}$, we have $f^{k,w}_a(c) = f^{w-k+1,w}_b(1-c)$ for all $c \in [0,1]$.
	\begin{proof}
		Recall that, by definition, $C_a$ and $C_b$ being symmetric implies $f_a(c) = f_b(1 - c)$ for all $c \in [0, 1]$, which, in turn, implies $F_a(c) = F_b(1 - c)$ for all $c \in [0, 1]$.
		We therefore have,
		\begin{align*}
			f^{k,w}_a(c) & = k  \binom{w}{k} f_a(c) [ F_a(c) ]^{k-1} [ 1 - F_a(c) ]^{w-k}\\
			& = k  \binom{w}{k} f_b(1-c) [1-F_b(1-c)]^{k-1} [ 1 - (1-F_b(1-c))]^{w-k}\\
			& = k  \frac{w!}{(w-k)!k!} f_b(1-c) [1-F_b(1-c)]^{k-1} [f_b(1-c) ]^{w-k}\\
			& = (w-k+1)\frac{w!}{(w-k+1)!(k-1)!} f_b(1-c) [ (1-F_b(1-c)]^{k-1} [F_b(1-c) ]^{w-k}\\
			& = (w-k+1)\frac{w!}{(w-k+1)!(w- (w - k + 1)!} f_b(1-c) [1-F_b(1-c)]^{k-1} [F_b(1-c) ]^{w-k}\\
			& = (w-k+1) 
			\binom{w}{w-k+1}
			%{w \choose w-k + 1} 
			f_b(1-c) [1-F_b(1-c)]^{k-1} [F_b(1-c) ]^{w-k}\\
			& = f^{w-k+1,w}_b(1-c)
		\end{align*} \end{proof}
\end{lem} 

%\begin{lem}
%\label{lem:order_one}
%Suppose that $f$ is symmetric.
%Then, for any $w \in \mathbb{N}$ and any $k \in \{1, \dots, w\}$, we have $f^{k,w}(x) = f^{w-k+1,w}(1-x)$ for all $x \in [0,1]$.
%%
%\begin{proof}
%%
%The result follows directly from Lemma \ref{lem:order} once we note that $C = C_a + C_b$ for two symmetric distribution pdfs $2\p_a$ and $2\p_b$ defined by $\p_a(x) = f(x)$ all for $x < 0.5$ (and $0$ otherwise), and $\p_b(x) = f(x)$ for all $x > 0.5$ (and $0$ otherwise).
%\end{proof}
%\end{lem}

%%%%%%%%%%%%%%%%%%%%%%%%%%%%%%%%%%%%
\subsection{Proof of Proposition \ref{prop:n_extreme}}
 %%%%%%%%%%%%%%%%%%%%%%%%%%%%%%%%%%%%
 
%We only prove the existence of $\ul{c}$ with the required property (the proof of the existence of $\bar{c}$ is symmetrical).
%Henceforth, we refer to jurors with conviction probability no larger than $c$ as \emph{extreme jurors}.
%Let $\ul{\Phi}^c_{M}(x)$ denote the probability that there are \emph{at least} $x$ extreme jurors in the jury selected by $M$.
%We prove the proposition by showing that for every $x \in \{0,\dots, j-1\}$, there exists $\ul{c}$ such that $\ul{\T}_{STR}(x ; c) < \ul{\T}_{\SAR}(x ; c)$ for all $c \in (0, \ul{c})$.

Consider an arbitrary $c \in (0,1)$ and let us refer to jurors with conviction probability no larger than $c$ as \emph{extreme jurors}.
Let $\ul{\T}_{M}(x ; c | k)$ denote the probability that at least $x$ extreme jurors are selected by procedure $M$ \emph{conditional} on there being exactly $k$ of extreme jurors in the panel of $n$.
By the Law of Total Probability,
    \begin{equation}
		\label{eq:decomp_more}
		\begin{split}
		\ul{\T}_{M}(x ; c) = \sum_{k = \green{x}}^n \p\Big[Bi\big(n, F(c)\big) = k\Big]~ \ul{\T}_{M}(x ; c | k). 
		\end{split}
		\end{equation}
	
	Consider first the \STR procedure. 
	Note that for all $c$, we have $\ul{\T}_{\STR}(x ; c | x) = 0$ because if there are exactly $x$ extreme jurors in the panel, one of them is necessarily challenged by the plaintiff under $\STR$ (recall that $p \geq 1$).
	Therefore, by \eqref{eq:decomp_more}, we have
		%		
	%	\begin{align}
	\begin{equation}
			\label{eq:comp_STR_more}
				\ul{\T}_{STR}(x ; c)    =  
				\sum_{k = \green{x+1}}^n \p\Big[Bi\big(n, F(c)\big) = k\Big] ~ \ul{\T}_{\STR}(x ; c | k)
				  \leq 
				 \p\Big[Bi\big(n, F(c)\big) > x \Big], 
	\end{equation}
	%	\end{align}
	where the last inequality follows from the fact that $\ul{\T}_{\STR}(x ; c | k) \in [0,1]$ for all $k$ (as $\ul{\T}_{\STR}(x ; c | k)$ is a probability).
		
		Next, consider procedure \SAR.
		Our goal is to construct a lower bound for the probability of selecting an extreme juror and show that, as $c \rightarrow 0$, this lower bound does not converge
        to 0 as fast as \eqref{eq:comp_STR_more}.
        To do so, we introduce a decreasing function $\sigma(c) > 0$  such that, when $c$ is sufficiently small, $\ul{\T}_{\SAR}(x ; c | k) \geq \sigma(c)$ for any $k \geq x$.
		To construct $\sigma$, consider the restricted sample space in which there are $k$ extreme jurors in the panel.
		
		Let $\ul{t}_P$ be the lowest challenge threshold used by the plaintiff in any subgame of \SAR.
		Clearly, $\ul{t}_P > 0$.\footnote{
		Formally, if $\Gamma$ denotes the set of subgames of \SAR and $t_P(\gamma)$ the plaintiff's challenge threshold in any subgame $\gamma \in \Gamma$, then $\ul{t}_P = \min_{\gamma \in \Gamma} t_p(\gamma)$ (the minimum is well-defined since $\Gamma$ is of finite size).
		In any subgame $\gamma$ of \SAR, there is always a $c > 0$ low enough such that if the juror who is presented to the parties in the first round of $\gamma$ is of type $c$, the plaintiff will challenge that juror.
		Therefore, $\ul{t}_P > 0$.
	}
		Henceforth, we focus on $c \in (0, \ul{t}_P)$.
		We first consider the function $\alpha(c)$ defined as the probability that $c_j \in  (c, \ul{t}_P)$ for \emph{all} the $(n - k)$ non-extreme jurors in the panel.
		Because $C$ is continuous and $0$ is the lower-bound of its support, there exists $y > 0$ sufficiently small such that $\alpha(c) > 0$ for all $c \in [0, y]$.\footnote{
			Because $0$ is the lower-bound of the defined support, $\p(C \in [0, \epsilon]) > 0$ for all $\epsilon > 0$.
			By continuity of $C$, there must therefore exists some $\delta > 0$ such that  $\p(C \in  [\delta/2 , \delta]) > 0$.
			We then have $\alpha(c) > 0$ for all $c < \delta$.
		} 
		Also, $\alpha(c)$ is weakly decreasing in $c$.
		By construction of $\ul{t}_P$, for such panels (with  $k$ extreme jurors and $c_j \in  (c, \ul{t}_P)$ for all the $(n-k)$ non-extreme jurors), the plaintiff uses all its challenges on the $p$ first jurors it is presented with, and the defendant never uses any challenges.\footnote{
		The latter follows because in any subgame the defendant's threshold is always higher plaintiff's (in equilibrium, the defendant and the plaintiff never both want to challenge the presented juror).}
		Hence, for these panels, the probability that all $k$ extreme jurors are selected is the probability that none of these jurors are among the $p$ first presented jurors, i.e., $\binom{n-p}{k}/\binom{n}{k}$.
		%${n-p \choose k}/{n \choose k}$.
		Overall, for $c \in (0, \ul{t}_P)$, we have $\ul{\T}_{\SAR}(x ; c | k) \geq \alpha(c) * \binom{n-p}{k}
		%{n-p \choose k}
		/ \binom{n}{k}
		%{n \choose k}
		$, and $\sigma(c) \coloneqq \alpha(c) * \binom{n-p}{k}/\binom{n}{k}$ has the desired property.
		
		Applying $\ul{\T}_{\SAR}(x ; c | k) \geq \sigma(c)$ to \eqref{eq:decomp_more} with $M = \SAR$, we obtain for all $c$ sufficiently small (specifically $c \in (0, \ul{t}_P)$)
		\begin{flalign}
			\label{eq:decomp_SAR_more}
				\ul{\T}_{\SAR}(x ; c)
				& \geq &
			\sum_{k = {x}}^n \p\Big[Bi\big(n, F(c)\big) = k\Big] ~*~ \sigma(c)
				& \geq &
				\p\Big[Bi\big(n, F(c)\big) = x\Big] ~*~ \sigma(c). 
		\end{flalign}

		Overall, combining  \eqref{eq:comp_STR_more} and \eqref{eq:decomp_SAR_more} yields
		\begin{align}
			\label{eq:final_ext}
			\lim_{c \ra 0}~ \frac{\ul{\T}_{\SAR}(x ; c)}{\ul{\T}_{STR}(x ; c)} 
			& \geq \lim_{c \ra 0}~ 
			\frac{\p\Big[Bi\big(n, F(c)\big) = x\Big]~ *~ \sigma(c)}
			{\p\Big[Bi\big(n, F(c)\big) > x \Big]}   = \infty,
		\end{align}
		where the last equality follows from Lemma \ref{lem:binom_gen} and the fact that 
		$\sigma(c) > 0$ is decreasing in $c$.\footnote{
		To apply Lemma \ref{lem:binom_gen}, note that because $C$ is continuous and the lower-bound of the support of $C$ is $0$, we have $F(c) > 0$ for all $c > 0$ and $\lim_{c \ra 0} F(c) = 0$. 
	}
		In turn, $\lim_{c \ra 0} {\ul{\T}_{\SAR}(x ; c)}/{\ul{\T}_{STR}(x ; c)} = \infty$ and the fact that $\lim_{c \ra 0} \ul{\T}_{\SAR}(x ; c) = \lim_{c \ra 0} \ul{\T}_{STR}(x ; c) = 0$ together  imply implies that there exists some $\ul{c} > 0$ small enough such that $\ul{\T}_{STR}(x ; c) < \ul{\T}_{\SAR}(x ; c)$ for all $c \in (0, \ul{c})$.
 
 %%%%%%%%%%%%%%%%%%%%%%%%%%%%%%%%%%%%
 \subsection{Proof of Proposition \ref{prop:extremeRAN}}
 %%%%%%%%%%%%%%%%%%%%%%%%%%%%%%%%%%%%
 
 Using the same notation as in the proof of Proposition \ref{prop:n_extreme}, we have
 \begin{align}
 	\label{eq:decomp_RAN_more}
 	\ul{\T}_{\RAN}(x ; c)
 	~~ \geq ~~ 
 	\p\Big[Bi\big(n, F(c)\big) = x\Big] ~*~ \ul{\T}_{\RAN}(x ; c| x). 
 \end{align}
 Note that $\ul{\T}_{\RAN}(x ; c| x)$ is the probability that an Hypergeometric random variable with $x$ success, $j-x$ failures, and $j$ draws, results in the draw of exactly $x$ successes.
 Therefore, $\ul{\T}_{\RAN}(x ; c| x) > 0$.
 Finally, combining \eqref{eq:decomp_RAN_more} and \eqref{eq:comp_STR_more} yields
 \begin{align*}
 	\lim_{c \ra 0}~ \frac{\ul{\T}_{\RAN}(x ; c)}{\ul{\T}_{STR}(x ; c)} 
 	& \geq \lim_{c \ra 0}~ 
 	\frac{\p\Big[Bi\big(n, F(c)\big) = x\Big]~ *~ \ul{\T}_{\RAN}(x ; c| x)}
 	{\p\Big[Bi\big(n, F(c)\big) > x \Big]}   = \infty,
 \end{align*}
 where the last equality follows from Lemma \ref{lem:binom_gen} and the fact that $\ul{\T}_{\RAN}(x ; c| x) > 0$.
 The result then follows as in the proof of Proposition \ref{prop:n_extreme}.

 %%%%%%%%%%%%%%%%%%%%%%%%%%%%%%%%%%%%
 \subsection{Proof of Proposition \ref{prop:minor_compare}}
 %%%%%%%%%%%%%%%%%%%%%%%%%%%%%%%%%%%%
 
  The structure of the proof is similar to that of the previous propositions. 
We focus on the case we analyzed in the main paper, where the minority uniformly favors the defendant, i.e., $\lim_{i \ra \infty} \p(C_a^i > C_b^i) = 0$.
 The proof for the other case is symmetrical.
 
 As in the previous proofs, for any arbitrary triple $(C^i_a, C^i_b, r^i)$, we first decompose $\M^i_{\STR}(x)$ and $\M^i_{\SAR}(x)$ by conditioning on the number of minority jurors in the panel.
 
 First, consider \STR and let us decompose $\M^i_{\STR}(x)$ conditional, on the one hand, on the panel containing more than $x$ minority jurors --- which occurs with probability $\p \big[Bi(n, r^i) >  x  \big]$, and on the other, on the panel containing exactly $x$ minority jurors --- which occurs with probability $\p \big[Bi(n, r^i) =  x  \big]$.
 In the first case (i.e., more than $x$ minority jurors in the panel),
 the probability that the panel contains at least $x$ minority jurors is an upper bound on the probability that \STR selects them.
 % assuming that \STR \emp{always} selects at least $x$ minority jurors clearly provides an upper-bound on the true probability that the latter occurs.
 In the second case (i.e., exactly $x$ minority jurors in the panel), \STR selects at least $x$ minority jurors provided that none of the minority jurors in the panel are challenged.
 This occurs with a probability no larger than the probability that the lowest conviction probability among minorities is larger than the $p$-th conviction probability among majority jurors (since the latter is required for the plaintiff not to challenge any of the minority jurors in the panel).
 Recall that for any number of draws $w$ and any $k\leq w$, we let $C^{k,w}_g$ denote the $k$-th order statistic out of $w$ draws from group $g \in \{a,b\}$.
 With this notation, we therefore have,
 \begin{align}
 	\label{eq:STR_minor}
 	\M^i_{\STR}(x) \leq \p \big[Bi(n, r^i) >  x  \big] + \p \big[Bi(n, r^i) =  x  \big] * \p\big([C_a^i]^{1,x} > [C_b^i]^{p,n-x}\big).
 \end{align}
 Note that because $\lim_{i \ra \infty} \p(C_a^i > C_b^i) = 0$, we have $\lim_{i \ra \infty} \p\big([C_a^i]^{1,x} > [C_b^i]^{p,n-x}\big) = 0$.	
 
 Second, consider \SAR.
 Clearly, $\M^i_{\SAR}(x)$ is no smaller than the probability for \SAR to select at least $x$ minority jurors when there are exactly $x$ minority jurors in the panel.
 The latter is equal to $\p \big[Bi(n, r^i) =  x  \big] * \sigma(x; r^i, C^i_a, C^i_b)$, where $\sigma(x; r^i, C^i_a, C^i_b)$ denotes the probability that \SAR selects $x$ minority jurors \emph{conditional} on having $x$ minority jurors in the panel, as a function of $r^i$, $C^i_a$, and $C^i_b$.
 In summary, with this notation, we have,
 \begin{align}
 	\label{eq:SAR_minor}
 	\M^i_{\SAR}(x) \geq  \p \big[Bi(n, r^i) =  x  \big] * \sigma(x; r^i, C^i_a, C^i_b).
 \end{align}

 We now show that $\lim_{i \ra \infty} \sigma(x; r^i, C^i_a, C^i_b) > 0$.
 For all $i \in \mathbb{N}$, let $C^i = r^i C^i_a + (1-r^i) C^i_b$.
 Observe that because $\lim_{i \ra \infty} r_i = 0$ and because $C^i_b$ converges in distribution to $C^*_b$, $C^i$ converges in distribution to $C^*_b$.
 By Lemma \ref{lem:converg}, this implies that for any subgame $\gamma$ of \SAR and both $I \in \{D,P\}$, we have $\lim_{i \ra \infty} {t}_I(\gamma, C^i\big) = {t}_I(\gamma, C^*_b\big)$.
 Note that ${t}_I(\gamma, C^*_b\big)$ lies in the interior of the support of $C^*_b$ for both $I \in \{D,P\}$.
 Also recall that in the limit, the supports of $C^i_a$ and $C^i_b$ do not overlap as we have $\p(C^*_a > C^*_b) = 0$.
 Therefore, in the limit, the defendant never challenges a minority juror, which in turn implies that 
 \begin{description}
 \item[(a)] as $i$ tends to infinity, the probability that the defendant challenges one of the $x$ minority jurors in the panel tends to zero.
 \end{description}
 Because ${t}_I(\gamma, C^*_b\big)$ lies in the interior of the support of $C^*_b$ for both $I \in \{D,P\}$, there is also a range of conviction probabilities $[\ul{c},\ol{c}]$ low enough inside the support of $C^*_b$ such that $P(C^*_b \in [\ul{c},\ol{c}]) > 0)$ and $P$ challenged the juror presented in subgame $\gamma$ if her conviction probability lies within $[\ul{c},\ol{c}]$.
 Furthermore, the probability that a juror with %conviction-probability in 
 $c\in [\ul{c},\ol{c}]$ is a majority juror is strictly positive (and tends to one as $i\rightarrow \infty$).
 Overall, in the limit,
 \begin{description}
 \item[(b)]the probability that the plaintiff challenges a majority juror presented in subgame $\gamma$ is strictly positive.
 \end{description}
 
 Combining (a) and (b), in the limit and given a panel containing $x$ minority jurors, there is a positive probability that $p$ majority jurors are presented first, are all challenged by $P$, and are followed by the $x$ minority jurors which are left unchallenged by the parties (resulting in a jury composed of at least $x$ minority jurors).
 That is, $\lim_{i \ra \infty} \sigma(x; r^i, C^i_a, C^i_b) > 0$.
 
 We are now equipped to complete the proof.
 Combining \eqref{eq:STR_minor} and \eqref{eq:SAR_minor} yields
 \begin{align*}
 	& \lim_{i \ra \infty} \frac{\M^i_{\STR}(x)}{\M^i_{\SAR}(x)} \\
 	\leq & \lim_{i \ra \infty} \frac{\p \big[Bi(n, r^i) >  x  \big] + \p \big[Bi(n, r^i) =  x  \big] * \p\big([C_a^i]^{1,x} > [C_b^i]^{p,n-x}\big)}{
 		\p \big[Bi(n, r^i) =  x  \big] * \sigma(r^i, C^i_a, C^i_b)}\\
 	= & \lim_{i \ra \infty} \frac{\p \big[Bi(n, r^i) >  x  \big]}{
 		\p \big[Bi(n, r^i) =  x  \big] * \sigma(r^i, C^i_a, C^i_b)} 
 	+
 	\frac{\p \big[Bi(n, r^i) =  x  \big] * \p\big([C_a^i]^{1,x} > [C_b^i]^{p,n-x}\big)}{
 		\p \big[Bi(n, r^i) =  x  \big] * \sigma(r^i, C^i_a, C^i_b)}\\
 	= & \lim_{i \ra \infty} \frac{\p \big[Bi(n, r^i) >  x  \big]}{
 		\p \big[Bi(n, r^i) =  x  \big]}
 	*
 	\frac{1}{\sigma(r^i, C^i_a, C^i_b)} 
 	+
 	\frac{\p\big([C_a^i]^{1,x} > [C_b^i]^{p,n-x}\big)}{
 		\sigma(r^i, C^i_a, C^i_b)} \\
 	= & \underbrace{\lim_{i \ra \infty} \frac{\p \big[Bi(n, r^i) >  x  \big]}{
 			\p \big[Bi(n, r^i) =  x  \big]}}_{ = 0, \text{ by Lemma \ref{lem:binom_gen}}}
 	* \underbrace{\lim_{i \ra \infty}
 		\frac{1}{\sigma(r^i, C^i_a, C^i_b)}}_{< \infty, \text{ by} \lim_{i \ra \infty} \sigma(x; r^i, C^i_a, C^i_b) > 0} 
 	+ \underbrace{\lim_{i \ra \infty}
 		\frac{\p\big([C_a^i]^{1,x} > [C_b^i]^{p,n-x}\big)}{
 			\sigma(r^i, C^i_a, C^i_b)}}_{
 		\substack{ =0,\\	 
 			\text{ by} \lim_{i \ra \infty} \p([C_a^i]^{1,x} > [C_b^i]^{p,n-x}) = 0,\\
 			\text{ and} \lim_{i \ra \infty} \sigma(x; r^i, C^i_a, C^i_b) > 0}} = 0 
 \end{align*}
 In turn, $\lim_{i \ra \infty} {\M^i_{\STR}(x)}/{\M^i_{\SAR}(x)} \leq 0$ and  $\lim_{i \ra \infty} {\M^i_{\STR}(x)} = \allowbreak \lim_{i \ra \infty} \allowbreak {\M^i_{\SAR}(x)} = 0$ together imply that $\exists {k}$ sufficiently large such that $\M^i_{\SAR}(x) > \M^i_{\STR}(x)$ for all  $i > k$.
 
 %%%%%%%%%%%%%%%%%%%%%%%%%%%%%%%%%%%%
 \subsection{Proof of Proposition \ref{prop:n_chall}}
 %%%%%%%%%%%%%%%%%%%%%%%%%%%%%%%%%%%%
 
 The structure of the proof is similar to that of the previous propositions.
 Observe that \eqref{eq:comp_STR_more} and \eqref{eq:decomp_SAR_more} are true regardless of the number of challenges awarded to the parties in \STR or \SAR.
 That is, by the same arguments as in the proof of Proposition \ref{prop:n_extreme}, the following two inequalities hold regardless of the values of $w$, $y$, $\M_{\STR\ph w}(x)$, or $\M_{\SAR\ph y}(x)$,\footnote{
 	Recall that the proposition assumes $w,y\geq 1$.}  
 \begin{align}
 	\label{eq:ext_n_chal}
 	\begin{split}
 		\ul{\T}_{\STR\ph w}(x ; c)    =  
 		\sum_{k = \green{x+1}}^n \p\Big[Bi\big(n, F(c)\big) = k\Big] ~ \ul{\T}_{\STR\ph w}(x ; c | k)
 		\leq 
 		\p\Big[Bi\big(n, F(c)\big) > x \Big],\\
 		\ul{\T}_{\SAR\ph y}(x ; c)
 		\geq 
 		\sum_{k = {x}}^n \p\Big[Bi\big(n, F(c)\big) = k\Big] ~*~ \sigma(c)
 		\geq 
 		\p\Big[Bi\big(n, F(c)\big) = x\Big] ~*~ \sigma(c). 
 	\end{split}
 \end{align}
The proof follows as in the proof of Proposition \ref{prop:n_extreme} (in particular, see \eqref{eq:final_ext}).

\subsection{Proof of proposition \ref{prop:statdisc}}

As in the baseline model, the plaintiff aims at maximizing (and the defense, minimizing) the expected probability of conviction \emph{conditional on the jurors' signals and group identity}, which in the statistical discrimination model is $ E(\Pi_{i=1}^jE(c_i|s_i, g_i)).$
Under \STR, after questioning parties will compute conditional expected probability of voting for conviction $\xi_i=E(c_i|s_i,g_i)$ and challenge the prospective jurors with the most extreme values of these conditional expectations. Hence, they behave as in a baseline model with $\xi_i$ equal to the true conviction probability.\footnote{The statement of the proposition refers only to equilibrium strategies. However, since values of $\xi_i$ are drawn from $q_{a}, q{_b}$ by members of groups $a,b$, it is also true that in each round of selection, outcomes of the statistical discrimination model are probabilistically equivalent to the outcomes of a baseline model with conviction probabilities drawn from $q_{a}, q{_b}$, and learned with certainty}

Under \SAR, we need to distinguish two cases. (i) in all subgames except those where only one challenge is left, parties will compare the conditional expected conviction probability of the presented juror $\xi_i$ with the value of waiting for the next round and being presented with a juror with unknown signal and group identity, from which the conditional expectation $\xi_i^{t+1}=E(c_i|s_i,g_i)$ is computed. From the parties' point of view, future realization  of the conditional expected probability of the member of a yet unknown group occurs with density $q(\xi_i^{t+1})=rq_a(\xi_i^{t+1})+(1-r)q_b(\xi_i^{t+1})$. 
The parties strategies are therefore identical to a baseline model in which the conviction probability of the current juror is $\xi_i$, and future values are drawn from $q_a$ with probability $r$, and from $q_b$ with probability $1-r$. (ii) In subgames with only one challenge left, if the challenge is exercised there is no signal extraction and learning in the following round(s). The parties know that the next juror's $c$ will be drawn from $f=rf_a+(1-r)f_b$. What matters in determining choices in these subgames is the expected value $E(f)$, which is equal to $E(rq_a+(1-r)q_b)$ by the law of iterated expectations, hence again in the statistical discrimination model parties behave as in a baseline model with conviction probabilities drawn from $q_a, q_b$, and group identity equal to $a,b$ with probability $r,1-r$. The proposition follows.

 \subsection{Proof of Proposition \ref{prop:median}}
 %%%%%%%%%%%%%%%%%%%%%%%%%%%%%%%%%%%%
 
 The probability that $\STR$ selects at least $x$ jurors with conviction probability above the median is the probability that at least $x + d$ of the jurors in the panel have conviction-probability above the median (since $d$ of these jurors are challenged by the defendant).
 Because $d = p$, for any $x \in \{1, \dots, j\}$, we therefore have
 \begin{align*}
     \ul{\T}_{\STR}\big(x ; med[C]\big) = P[Bi(j + d + p,0.5) \geq x + d] = P[Bi(j + 2d,0.5) \geq x + d]
 \end{align*}
 In contrast, we have
 \begin{align*}
     \ul{\T}_{\RAN}\big(x ; med[C]\big) & = P[Bi(j,0.5) \geq x],
 \end{align*}
therefore, by repeated application of Lemma \ref{lem:comp_stat}, $x > (j/2) + (1/2)$ implies $\ul{\T}_{\STR}\big(x ; med[C]\big) > \ul{\T}_{\RAN}\big(x ; med[C]\big)$.
Since $j$ is integer-valued, the last inequality corresponds to $x \geq j/2 + 1$ if $j$ is even and $x \geq j/2 + 1.5$ if $j$ is odd.

%%%%%%%%%%%%%%%%%%%%%%%%%%%%%%%%%%%%
\subsection{Proof of Proposition \ref{prop:compare_sym}}
%%%%%%%%%%%%%%%%%%%%%%%%%%%%%%%%%%%%

%%%%%%%%
\textbf{Part (a).}
Under \STR, since the group-distributions do not overlap, each party first uses all of its challenges on one of the two groups before challenging the lowest conviction probability jurors from the other group.
For concreteness and without loss of generality, suppose that group $a$ favors the defendant (i.e., $\p(C_a > C_b) = 0$).
Let $m$ denote the number of jurors from group-$a$ in the panel.

Note that because $r=0.5$, the probability that $m = k$ is the same as the probability that $m = n - k$ for all $k \in \{1, \dots, \floor{n/2}\}$.
% \footnote{
% This follows from the symmetry of the binomial distribution when the probability of success is 0.5 ($\p[m = i] = P[Bi(n,0.5) = i] = P[Bi(n,0.5) = z-i] = \p[m = z-i]$ since ${z \choose i} = {z \choose z-i}$).
% }
Also, because $d=p$, the number of group-$a$ jurors who are selected when $m=k$ is equal to the number of group-$b$ jurors who are selected when $m = n - k$.\footnote{
First, suppose that $k \leq p$.
Then, if $m = k$, no jurors from group-$a$ (and $j$ jurors from group-$b$) are selected,
whereas if $m = n - k$, no jurors from group-$b$ (and $j$ jurors from group-$a$) are selected.
Second, suppose that $k \in \{p+1, \dots, \floor{n/2} \}$.
Then, if $m = k$, $k-p = k-d$ jurors from group-$a$ (and $j - (k-p) = j - (k-d)$ jurors from group-$a$) are selected, 
whereas if $m = n - k$, $k-d = k-p$ jurors from group-$b$ (and $j - (k-d) = j - (k-p)$ jurors from group-$b$) are selected. 
}
Therefore, the expected number of group-$a$ jurors in the jury selected by \STR is exactly $j/2$.
% \footnote{
% Formally, let $(\#m|i)$ denote the number of group 0 jurors who are selected conditional on the panel containing $i$ group 0 jurors. 
% If $n$ is even, we then have
% %
% \begin{align*}
%     r_{STR} & =  \sum_{i = 1}^{\floor{n/2}} P(m = i)*(\#m|i) + P(m = n-i)*(\#m|n-i)\\
%             & =  \sum_{i = 1}^{\floor{n/2}} P(m = i)*(\#m|i) + P(m = n-i)*[j - (\#m|i)] \\
%             & =  \sum_{i = 1}^{\floor{n/2}} P(m = i)*(\#m|i) + P(m = i)*[j - (\#m|i)] \\
%             & =  \sum_{i = 1}^{\floor{n/2}} P(m = i)*(\#m|i)  - \sum_{i = 1}^{\floor{n/2}} P(m = i)*(\#m|i) + \sum_{i = 1}^{\floor{n/2}} P(m = i)*j \\
%             & = j*\sum_{i = 1}^{\floor{n/2}} P(m = i) = j*/2
% \end{align*}
% %
% The argument is similar when $n$ is odd.
% }

%%%%%%%%
\medskip
\textbf{Part (b).}
The proof is similar to the proof of Part (a). 
Consider the set of panel configurations $\{a,b\}^n$ where, for example, vector $(a,b,a,\dots,b,b,b) \in \{a,b\}^n$ indicates that the juror with the lowest conviction probability in the panel is a group-$a$ juror, the juror with second-lowest conviction probability is a group-$b$ juror, the juror with the third-lowest conviction probability is a group-$a$ juror, ..., and the jurors with the three highest conviction probabilities are all group-$b$ jurors.
To explain the structure of the proof, suppose that $n$ is even (we explain below how the argument generalizes to any $n$).
We first construct a partition of $\{a,b\}^n$ into two subsets $S^a$ and $S^b$ of equal size and construct a bijection $q$ between $S^a$ and $S^b$.
We then show that for every panel configuration $l \in S^a$ which results in $m^l$ group-$a$ jurors being selected, 
(a) the panel configuration $q[l]$ result $j - m^l$ group-$a$ jurors being selected, and
(b) panel configurations $l$ and $q[l]$ are equally likely.
As in the proof of Part (a), the result then follows directly.

Similar to the proof of Part (a), the bijection $q[l]$ is obtained by
(i) mirroring $l$ around the $\floor{n/2}$ position, and
(ii) inverting the group of each juror in the resulting panel configuration.
For example, panel configuration $q[(a,a,b,a)]$ is obtained by mirroring $(a,a,b,a)$ around position $\floor{n/2}$, which results in $(a,b,a,a)$, and then inverting the group of each jurors in $(a,b,a,a)$, which results in $(b,a,b,b)$.
Formally, if $inv[l]$ denotes the configuration that results from turning all the $a$'s in $l$ into $b$'s and all the $b$'s in $l$ into $a$'s, then  $q[(l_1, l_2, \dots, l_{n-1}, l_n)] = inv[(l_n, l_{n-1}, \dots, l_2, l_1)]$.

Let $S^a$ and $S^b$ be two sets that together contain all $l$ for which $l \neq q[l]$ and are such that $l\in S^i$ implies $q[l] \notin S^i$.
Since $q\big[ q[l] \big] = l$, the sets $S^a$ and $S^b$ have equal sizes.
Also let $S^*$ contain all $l$ for which $l = q[l]$, if any ($S^* \neq \emptyset$ if and only if $n$ is even).
Note that $\{S^a, S^b, S^*\}$ forms of partition of $\{a,b\}^n$.
Therefore, if we let $(\#m|l)$ denote the number of group-$a$ juror that are selected conditional on configuration $l$ and $\p(l)$ the probability of configuration $l$, we have
\begin{align*}
    r_{STR} = \sum_{l \in S^a} \p(l)*(\#m|l) + \p(q[l])*(\#m|q[l]) + \sum_{l \in S^*} \p(l)*(\#m|l).
\end{align*}
Part (b) then follows from the fact that 
(A) $\p(l) = \p(q[l])$ for all $l \in S^a$,
(B) $(\#m|l) = n - (\#m|q[l])$ for all $l \in S^a$, and
(C) $(\#m|l) = j/2$ for all $l \in S^*$.

Properties (B) and (C) follow directly from the construction of $q$ and the fact that $d = p$.
Property (A), on the other hand, follows from Lemma \ref{lem:order} which establishes the symmetry of order statistics for symmetric distributions.
A formal proof of (A) using Lemma \ref{lem:order} requires heavy and tedious notation.
Instead, we show how (A) follows from Lemma \ref{lem:order} in a simple example that clarifies how the argument generalizes to other cases. 

Consider the case of $(a,a,b)$ for which $q[(a,a,b)] = (a,b,b)$.
We can obtain the probability of any configuration by integrating the p.d.f. of the appropriate order statistics from the bottom to the top of $[0,1]$.
For example, using the notation for order statistics introduced before Lemma \ref{lem:order}, we have
\begin{align}
    \label{eq:order_1}
    \p[(a,a,b)] & = \p[m = 2] * P[(a,a,b) | m=2] \notag\\
    & = \p[Bi(3,0.5) = 2] * \int_a^1 f_a^{1,2}(x) \left[ \int_x^1 f_a^{2,2}(y) \left( \int_y^1 f_b^{1,1}(w) ~dw \right) ~dy \right] ~dx .
\end{align}
We can also obtain the probability of any configuration by reverting the list of order statistics and integrating from the top to the bottom of $[0,1]$.
For example,
\begin{align}
    \label{eq:order_2}
    & \p[(a,b,b)] \notag\\ 
    & = \p[m = 1] * P[(a,b,b) | m=1] \notag\\
    & = \p[Bi(3,a.5) = 1] * \int_a^1 f_b^{2,2}(1-x) \left[ \int_x^1 f_b^{1,2}(1-y) \left( \int_y^1 f_a^{1,1}(1-w) ~dw \right) ~dy \right] ~dx .
\end{align}
Finally, by Lemma \ref{lem:order}, $f_a^{1,2}(x) = f_b^{2,2}(1-x)$, $f_a^{2,2}(y) = f_b^{1,2}(1-y)$, and $f_b^{1,1}(w) = f_a^{1,1}(1-w)$, which together with symmetry of the binomial with 0.5 probability of success implies that the expressions in \eqref{eq:order_1} and \eqref{eq:order_2} are equal.

\subsection{Statistical discrimination model: posteriors derivation and additional simulations}\label{sec:posteriors}

\subsubsection{Logit-normal priors}

Random variable $c$ is distributed according to a Logit-normal distribution when $c=e^{x}/(1+e^{x})$ with $x\sim N(\mu,\sigma^2).$ Its support is $[0,1]$ and its density is 
$$
c\sim \frac{1}{c(1-c)\sigma\sqrt{2\pi}}e^{-\frac{1}{2}\left(\frac{Ln(\frac{c}{1-c})-\mu)}{\sigma}\right)^2}
$$

Consider the underlying auxiliary normal random variable $x$, and assume parties observe a noisy i.i.d. signal of $x$, $s=x+\epsilon$, with $\epsilon\sim(0,\sigma_\epsilon^2)$. Then from the properties of the bivariate normal,  $f(x|s)\sim 
N\left(\alpha s + (1-\alpha) \mu,
       \alpha\sigma_\epsilon^2\right)$ 
with $\alpha=\frac{\sigma^2}{\sigma^2+\sigma_\epsilon^2} $. 

Note that $s$ itself is distributed $\sim N(\mu, \sqrt{\sigma^2+\sigma_\epsilon})$.
Applying the transformation of variables rule to $\xi(s)=E(x|s)=\alpha s + (1-\alpha) \mu$, the distribution of the conditional expectation is 
$
q(z)\sim  N\left(\mu,\alpha\sqrt{\sigma^2+\sigma_\epsilon^2}\right).
$ 
Note that as $\sigma_\epsilon$ increases the variance of $q(z)$ is decreasing and the distribution concentrates around $\mu$ because the term $\alpha$ converges to zero. $E(c|s)$ displays a logit-normal distribution with the corresponding parameters.

\subsubsection{Beta priors and binomial signals}

Jurors from group $g=a,b$ draw their conviction probability $c$ from distribution
$Beta(\alpha_g,\beta_g)$. Parties observe group identity $g$ and $N$ binary signals  $s_j\in\{0,1\}, j=1\ldots N$ independently drawn with $\Pr(s_j=1)=c.$ 

We treat $N$ as a parameter indicating the signal's precision. The distribution of ``successes" among the N Signals is a Binomial with parameters ($N,c$). $N=0$ corresponds to the case where the only information available is group identity, whereas as As $N\rightarrow\infty$ for the law of large numbers, the fraction of successes converges to $c$. 

This parameterization is convenient because, borrowing well-known results from the Bayesian inference literature, the Beta distribution is the conjugate prior probability distribution for the Binomial distribution, and by Bayes' rule, denoting with $s$ the number of successes ($=\Sigma_j s_j$),
$
	f(c_i| s_1\ldots s_N, g) \sim Beta(\alpha_g+s, \beta_g+N-s).
$
Recall that the expected value of distribution $Beta(\alpha,\beta)=\alpha/(\alpha+\beta)$, therefore
$$
E(f(c_i|s_1\ldots s_N,g)) = \frac{\alpha_g+s}{\alpha_g+\beta_g+N}.
$$
As $N\rightarrow\infty$ for the law of large numbers the number of successes $s$ converges to $Nc_i$, this expected value converges to $c_i$, the true value.

\begin{figure}[t]
    \centering
    \caption{Minority representation in the Statistical Discrimination model with Beta-binomial distributions}\label{fig:beta-binomial}
    \includegraphics[width=4.5in]{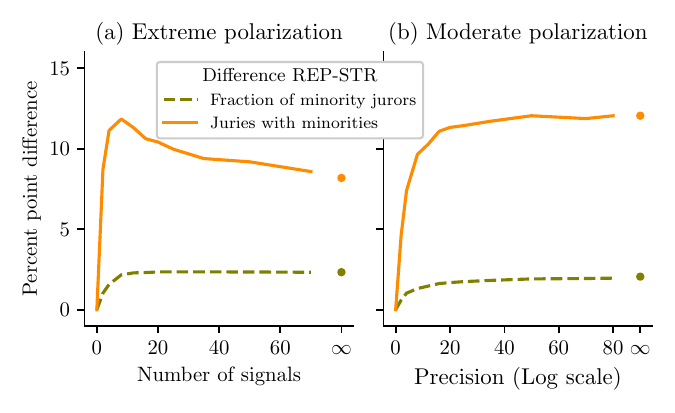}  
%    \caption*{\normalfont\footnotesize \emph{Notes:}  Percent points differences between \SAR and \STR in the number of seated jurors from group $a$ (dashed lines) and in juries with at least one juror from group $a$ (solid lines), $r=0.25$, 50000 simulated juries per simulation. Panel (a):  $f_a=Beta(1,5),f_b=Beta(5,1)$. Panel (b): $f_a=Beta(2,5),f_b=Beta(5,2).$ Signals in the horizontal axes are the number of Bernoulli trials with parameter $p=$ true conviction probability. Values corresponding to tick marker $\infty$ are computed using the baseline model. Each panel's inset displays the true distribution of conviction probabilities (brown solid line), each group's distributions (dashed lines), and the group averages (dotted vertical lines)
%    }
    
\end{figure}

The marginal probability of observing $x$ successes is 
\begin{align*}
p(x)= &\int_{0}^{1}f(x|c)f(c)dc=
  \int_{0}^{1}{x \choose N}c^{x}(1-c)^{N-x}\frac{1}{B(\alpha,\beta)}c^{\alpha-1}(1-c)^{\beta-1}dc=\\
 & \int_{0}^{1}{x \choose N}\frac{1}{B(\alpha,\beta)}c^{\alpha+x-1}(1-c)^{\beta+N-x-1}dc=\\
 & {x \choose N}\frac{B(\alpha+x,\beta+N-x)}{B(\alpha,\beta)}\int_{0}^{1}\frac{1}{B(\alpha+x,\beta+N-x)}c^{\alpha+x-1}(1-c)^{\beta+N-x-1}dc
\end{align*}
 where $B()$ is Euler's Beta function. Since the integrand is the beta distribution with parameters
$(\alpha+x,\beta+N-x)$, it integrates to 1. Therefore:
\[
p(s)={x \choose N}\frac{B(\alpha+x,\beta+N-x)}{B(\alpha,\beta)}
\]
and given that $E(c|s)$ is injective let $E^{-1}(z)$ be its inverse, then the distribution of the conditional expectations is $q(z)=p(E^{-1}(z))$
The domain $\Xi_g$ of $q_g$ is the set of rational numbers defined by the sequence 
$\{
\frac{\alpha_g+s}{\alpha_g+\beta_g+N}
\}_{s=0\ldots N}$.
    Figure \ref{fig:beta-binomial} displays results from simulations obtained from two such parameterizations, with results similar to those obtained in the simulations displayed in Section \ref{sec:discrimination}. In panel (a) we used $f_a=Beta(1,5),f_b=Beta(5,1)$, whereas in Panel (b) we used $f_a=Beta(2,5),f_b=Beta(5,2)$, with fraction of minorities in the pool  $r=0.25$.

%%%%%%%%%%%%%%%%%%%%%%%%%%%%%%%%%%%%
%%%%%%%%%%%%%%%%%%%%%%%%%%%%%%%%%%%%
%%%%%%%%%%%%%%%%%%%%%%%%%%%%%%%%%%%%
 
\bibliographystyle{aer}
\setlength{\bibsep}{0pt}  
\bibliography{library}

\end{document}